\documentstyle[12pt,amssymb,epsf]{article}

\makeatletter
\@addtoreset{equation}{section}
\makeatother

\def\ben{\begin{equation}}
\def\een{\end{equation}}

\let\a=\alpha    \let\e=\epsilon
    
 \let\m=\mu \let\n=\nu

\let\C=\Chi

\def\nn{\nonumber} \def\bd{\begin{document}} \def\ed{\end{document}}
\def\ds{\documentstyle} \let\fr=\frac \let\bl=\bigl \let\br=\bigr
\let\Br=\Bigr \let\Bl=\Bigl
\let\bm=\bibitem
\let\na=\nabla
\let\pa=\partial \let\ov=\overline
\newcommand{\be}{\begin{equation}}
\newcommand{\ee}{\end{equation}}
\def\ba{\begin{array}}
\def\ea{\end{array}}
\def\ft#1#2{{\textstyle{{\scriptstyle #1}\over {\scriptstyle #2}}}}
\def\fft#1#2{{#1 \over #2}}
\def\del{\partial}
\def\vp{\varphi}
\def\sst#1{{\scriptscriptstyle #1}}
\def\oneone{\rlap 1\mkern4mu{\rm l}}
\def\td{\tilde}
\def\wtd{\widetilde}
\def\ie{\rm i.e.\ }
\def\dalemb#1#2{{\vbox{\hrule height .#2pt
        \hbox{\vrule width.#2pt height#1pt \kern#1pt
                \vrule width.#2pt}
        \hrule height.#2pt}}}
\def\square{\mathord{\dalemb{6.8}{7}\hbox{\hskip1pt}}}
\newcommand{\ho}[1]{$\, ^{#1}$}
\newcommand{\hoch}[1]{$\, ^{#1}$}
\newcommand{\bea}{\begin{eqnarray}}
\newcommand{\eea}{\end{eqnarray}}
\newcommand{\ra}{\rightarrow}
\newcommand{\lra}{\longrightarrow}
\newcommand{\Lra}{\Leftrightarrow}
\newcommand{\ap}{\alpha^\prime}
\newcommand{\bp}{\tilde \beta^\prime}
\newcommand{\tr}{{\rm tr} }
\newcommand{\Tr}{{\rm Tr} }
\def\0{{\sst{(0)}}}
\def\1{{\sst{(1)}}}
\def\2{{\sst{(2)}}}
\def\3{{\sst{(3)}}}
\def\4{{\sst{(4)}}}
\def\5{{\sst{(5)}}}
\def\6{{\sst{(6)}}}
\def\7{{\sst{(7)}}}
\def\8{{\sst{(8)}}}
\def\n{{\sst{(n)}}}
\def\cA{{{\cal A}}}
\def\cB{{{\cal B}}}
\def\cF{{{\cal F}}}
\def\tV{\widetilde V}
\def\tW{\widetilde W}
\def\tH{\widetilde H}
\def\tE{\widetilde E}
\def\tF{\widetilde F}
\def\tA{\widetilde A}
\def\im{{{\rm i}}}
\def\tY{{{\wtd Y}}}
\def\ep{{\epsilon}}
\def\vep{{\varepsilon}}
\def\R{\rlap{\rm I}\mkern3mu{\rm R}}
\def\bD{{{\bar D}}}

\def\R{\rlap{\rm I}\mkern3mu{\rm R}}
\def\bD{{{\bar D}}}
\def\R{{{\Bbb R}}}
\def\C{{{\Bbb C}}}
\def\H{{{\Bbb H}}}
\def\CP{{{\Bbb C}{\Bbb P}}}
\def\RP{{{\Bbb R}{\Bbb P}}}
\def\Z{{{\Bbb Z}}}
\def\bA{{{\Bbb A}}}
\def\bB{{{\Bbb B}}}
\def\bC{{{\Bbb C}}}
\def\bD{{{\Bbb D}}}
\def\bE{{{\Bbb E}}}
\def\bZ{{{\Bbb Z}}}
\def\Re{{{\frak{Re}}}}
\def\Im{{{\frak{Im}}}}
\def\cosec{{\,\hbox{cosec}\,}}
\def\Gm{{\Gamma_{\!\! -}}}
\def\Gp{{\Gamma_{\!\! +}}}
\def\stan{{standard }}
\def\nonstan{{supernumerary }}

\newcommand{\tamphys}{\it Center for Theoretical Physics,
Texas A\&M University, College Station, TX 77843}

\newcommand{\upenn}{\it Department of Physics and Astronomy,\\ University
of Pennsylvania, Philadelphia, PA 19104}

\newcommand{\brussels}{\it Physique Th\'eorique et Math\'ematique,
Universit\'e Libre de Bruxelles,\\ Campus Plaine C.P. 231, B-1050
Bruxelles, Belgium}

\newcommand{\auth}{S. Cucu\hoch{\ast1}, H. L\"u\hoch{\dagger2} and J.F.
V\'azquez-Poritz\hoch{\ddagger3}}

\begin{document}
\begin{flushright}

KUL-TF-03-06\ \ \ \ \ MIFP-03-08\ \ \ \ \
ULB-TH/03-11\\
April  2003\ \ \ \ \ {\bf hep-th/0304022}\\
\end{flushright}


\begin{center}

{\large {\bf Interpolating from AdS$_{D-2}\times S^2$ to AdS$_D$ }}

\vspace{10pt}
\auth

\vspace{5pt}
{\hoch{\ast}\it Instituut voor Theoretische Fysica, Katholieke
Universiteit Leuven,\\ Celestijnenlaan 200D B-3001 Leuven, Belgium}

\vspace{5pt}
{\hoch{\dagger}\it George P. and Cynthia W. Mitchell Institute for
Fundamental Physics,\\ Texas A\& M University, College Station, TX
77843-4242, USA}

\vspace{5pt}
{\hoch{\ddagger}\brussels}


\underline{ABSTRACT}
\end{center}

      We investigate a large class of supersymmetric magnetic brane
solutions supported by $U(1)$ gauge fields in AdS gauged
supergravities.  We obtain first-order equations in terms of a
superpotential. In particular, we find systems which interpolate
between AdS$_{D-2}\times \Omega^2$ (where $\Omega^2=S^2$ or $H^2$)
at the horizon and AdS$_D$-type geometry in the asymptotic region,
for $4\le D\le 7$. The boundary geometry of the AdS$_D$-type
metric is Minkowski$_{D-3}\times \Omega^2$. This provides smooth
supergravity solutions for which the boundary of the AdS spacetime
compactifies spontaneously.  These solutions indicate the
existence of a large class of superconformal field theories in
diverse dimensions whose renormalization group flow runs from the
UV to the IR fixed-point.  We show that the same set of
first-order equations also admits solutions which are
asymptotically AdS$_{D-2}\times \Omega^2$ but singular at small
distance. This implies that the stationary AdS$_{D-2}\times
\Omega^2$ solutions typically lie on the inflection points of the
modulus space.

{\vfill\leftline{}\vfill \vskip 10pt \footnoterule {\footnotesize
\hoch{1} Research supported in part by the Federal Office for
Scientific, Technical and Cultural Affairs \phantom{of the}
through the Interuniversity Attraction Pole P5/27 and the European
Community's Human \phantom{of the} Potential Programme under
contract HPRN-CT-2000-00131 Quantum Spacetime.

{\footnotesize \hoch{2}
Research supported in part by DOE grant DE-FG03-95ER40917.

{\footnotesize \hoch{3} Research supported in part by the Francqui
Foundation (Belgium), the Actions de Recherche \phantom{of the}
Concert{\'e}es of the Direction de la Recherche Scientifique -
Communaut\'e Francaise de Belgique, \phantom{of the} IISN-Belgium
(convention 4.4505.86) and by a ``Pole d'Attraction
Interuniversitaire.''}} \vskip -12pt} } \pagebreak
\setcounter{page}{1}

\tableofcontents
\addtocontents{toc}{\protect\setcounter{tocdepth}{2}}
\newpage

\section{Introduction}

        Anti-de Sitter (AdS) spacetime has a natural boundary which,
under the Poincar\'e patch, is a Minkowski spacetime.  This provides a
simple but non-trivial example with which the holographic principle
can be tested through the AdS/CFT correspondence
\cite{malda,gkp,wit}. Since AdS spacetime can exist in
higher-dimensions ($D_{\rm max}=7$ in supergravities), it is worth
studying the possibility of spontaneous compactification of the AdS
boundary.  Any Ricci-flat perturbation on the boundary is consistent
with the equations of motion of pure Einstein gravity with a negative
cosmological constant. It is of interest to obtain solutions in gauged
supergravities whose asymptotic geometry is described by the metric
\be d\hat s^2_D = e^{2k z}\, ds_{D-1}^2 + dz^2\,,\label{adslike}
\ee
where the metric $ds_{D-1}^2$ is not Ricci-flat.  This metric cannot
be Einstein for general $z$, since the corresponding Ricci-tensor is
given by
\bea
\hat R_{ab} &=& e^{-2k z}\, R_{ab} - (D-1)\, k^2\, \eta_{ab}\nn\\
\hat R_{zz} &=& - (D-1)\, k^2\,.
\eea
However, at the boundary with $z\rightarrow \infty$, the spacetime
is Einstein even for non-vanishing $R_{ab}$, provided that it is
smooth. In this paper, we consider boundary metrics of the type
\be
ds_{D-1}^2 = dx^\mu\, dx_\mu + \lambda^{-2}\, d\Omega_2^2\,,
\label{boundmetric}
\ee
where $d\Omega_2^2$ is the metric of a two-sphere $S^2$, hyperbolic
two-plane $H^2$ or two-torus $T^2$. We will refer to the metric
(\ref{adslike}), with the boundary given by (\ref{boundmetric}), as
the AdS-type metric.

    Brane solutions whose boundary metric is given by
(\ref{boundmetric}) have been studied in various dimensions
\cite{romans, klemm1, sabra3, klemm2, sabra4, nunez, nunez2,
gauntlett4, sabra1, sabra2, recent}. They are all supported by a
set of 2-form field strengths, which are proportional to the
volume of $\Omega^2$, together with a set of scalars and a
superpotential. Thus, these solutions can be viewed as the
magnetic duals of the AdS black hole solutions obtained in
\cite{bh1,bh2,bh3,bh4,bh5,tenauthors}. Smooth and supersymmetric
solutions of this type have hitherto been limited to $H^2$, with
the exception of a smeared NS5-brane on $S^2$ \cite{nunez2}.

        The case of $S^2$, with the radius given by $\lambda^{-1}$,
provides a simple procedure for wrapping extra dimensions. This
enables one to make contact between $D$-dimensional gauged
supergravities and $(D-3)$-dimensional quantum field theories. In
fact, investigations into the possibility that our $D=4$ world arises
as a spontaneous compactification of a $D=6$ theory on
Minkowski$_4\times S^2$ has a long history . In 1953, Pauli attempted,
in one of his many unpublished works, to obtain an $SU(2)$ Yang-Mills
field from such a compactification of pure six-dimensional Einstein
gravity \cite{pauli}. Such a solution was later obtained in \cite{cj},
where additional gauge and Higgs fields, together with a scalar
potential, were added. A six-dimensional supergravity theory with such
a solution was constructed in \cite{sezgind6}. However, the string or
M-theory origin of this supergravity theory remained to be
understood. In this paper, we have obtained brane solutions in $D=7$
gauged supergravity whose boundary is precisely the metric
(\ref{boundmetric}).  This suggests that the effective action at the
boundary (with an UV cut-off for inducing gravity) of our brane
solution is precisely the aforementioned six-dimensional gauged
supergravity.

         We use a simple method to obtain first-order equations which
describe such solutions and obtain smooth solutions for which
$d\Omega_2^2$ is the metric for an $S^2$, as well as an $H^2$ and
$T^2$. We begin by performing a Kaluza-Klein reduction on $S^2$,
keeping only the singlet of the group action $SO(3)$. The resulting
$(D-2)$-dimensional theory consists of the metric and a set of scalars
with a scalar potential. We find that there exists a superpotential,
which enables us to obtain the first-order equations for a domain wall
solution. Lifting this system back to $D$ dimensions yields a magnetic
$(D-3)$-brane.

       With this method, we obtain the first-order equations for a
large class of solutions, which include previously-known solutions as
well as new ones. Although we have not found the most general
analytical solutions, their qualitative structures can easily be
investigated. In particular, numerical analysis shows that there is a
large class of solutions which smoothly interpolate between AdS$_{D-2}
\times \Omega^2$ geometry in the horizon and AdS$_D$-type geometry in
the asymptotic region, where $\Omega^2$ can be $S^2$ or $H^2$ and
$4\le D\le 7 $. There are also solutions which have the asymptotic
geometry AdS$_{D-2}\times \Omega^2$ and are singular at small
distance. The stationary solutions of AdS$_{D-2}\times \Omega^2$ with
constant scalars and their M-theory and string theory interpretation
have already been reported in a letter \cite{adsletter}.

         This paper is organized as the follows.  In the next section,
we re-derive magnetic brane solutions in $D$-dimensional AdS
Einstein-Maxwell gravity by using the $(D-2)$-dimensional
superpotential. In sections 3 and 4, we use the superpotential
method to find single-charge and two-equal-charge branes in
$D$-dimensional AdS gauged supergravities, respectively. In
sections 5-8, we consider general multi-charge branes in various
dimensions. Analytical solutions are found for the single-charge
branes, whereas numerical analysis is used when there are multiple
charges. The ten and eleven dimensional origins of our solutions
are described in section 9. Conclusions are presented in section
10.

\section{Branes in AdS Einstein-Maxwell gravity}

         Let us consider a $D$-dimensional theory with the Lagrangian
\be
\hat e^{-1}\, \hat{\cal L}= \hat R - \fft{1}{2n!}\, \hat F_\n^2 +
\Lambda\,,
\ee
where $\hat e=\sqrt{-\hat g}$, $\hat F_\n=d\hat A_{\sst{(n-1)}}$
and the cosmological constant is given by $-\Lambda$. We perform a
dimensional reduction on an $n$-dimensional space with the ansatz
\bea
d\hat s_{D}^2 &=& e^{2\alpha\, \varphi}\, ds_{D-n}^2 +
\lambda^{-2}\, e^{2\beta\, \varphi}\, d\Omega_n^2\,,\nn\\
\hat F_\n&=&\epsilon\, m\,\lambda^{-n}\,\Omega_\n\,, \label{reduce1}
\eea
where
\be
\alpha = -\sqrt{\fft{n}{2(D-2)(D-n-2)}}\,, \qquad
\beta = -\fft{(D-n-2)\,\alpha}{n}\,. \label{reduce2}
\ee
The reduction is consistent and the resulting $(D-n)$-dimensional
Lagrangian becomes
\be e^{-1}\, {\cal L} = R - \ft12 (\del\varphi)^2 - V(\varphi)\,,
\ee
where the scalar potential is given by \cite{instanton}
\be
V= \ft12 \epsilon^2 m^2 e^{2(D-n-1)\,\alpha\,\varphi} -
\epsilon\,n(n-1)\,
\lambda^2\, e^{\ft{2(D-2)}{n}\alpha\, \varphi}-\Lambda\,
e^{2\alpha\,\varphi}\,.\label{nondilsp}
\ee
The constant $\epsilon =1$, $-1$ or $0$, corresponding to
$d\Omega_n^2$ being the metric for a unit $n$-sphere, hyperbolic
$n$-plane or $n$-torus, respectively. We would like to express the
scalar potential (\ref{nondilsp}) in terms of a superpotential
$W$. Namely,
\be
V=(\fft{\del W}{\del \varphi})^2 - \fft{D-n-1}{2(D-n-2)}\, W^2\,.
\ee
We find that this to be possible only for $n=2$, in which case the
superpotential is given by
\be
W=\sqrt{\fft{D-2}{D-3}} (\epsilon\, m\, e^{(D-3)\,\alpha\, \varphi} +
2 \lambda^2\, m^{-1}\, e^{\alpha\, \varphi})\,, \label{W}
\ee
provided that the constraint $\Lambda=\fft{2(D-1)\, \lambda^4}{(D-3)\,
m^2}$ is satisfied.  Note that, in the case of $\epsilon=0$ which
corresponds to a 2-torus, the 2-form field strength vanishes. In fact,
in this case we find that, if we do turn on $F_\2$, then there is a
superpotential only in eleven dimensions, which is given by
\be
W=\fft{1}{2\sqrt2} \Big(3m\, e^{-\ft{8}{3\sqrt7}\, \varphi} +
\Lambda\, m^{-1}\, e^{\ft2{\sqrt7}\, \varphi}\Big)\,.
\ee
Though a curious feature, since there is no vector field in the
eleven-dimensional sector of M-theory we shall not further
consider this case.

    We can now construct a domain wall solution in $(D-2)$ dimensions,
with the metric ansatz
\be ds_{D-2}^2 = e^{2A}\, dx^{\mu}\, dx_\mu+ dy^2\,,
\ee
where $A$ and $\varphi$ are assumed to depend only on the
transverse coordinate $y$. The equations of motion are given by
\bea
&&\varphi'' + (D-3)\, A'\, \varphi' = \fft{\del V}{\del
\varphi}\,, \qquad A'' + (D-3)\, A'^2 = -\fft{V}{D-4}\,,\nn\\
&&\ft12 \varphi'^2 - (D-3)\, (D-4)\, A'^2 = V\,,
\eea
where a prime denotes a derivative with respect to $y$. Since the
scalar potential can be expressed in terms of a superpotential,
these second-order equations can be solved by the first-order
equations
\be
\varphi'=\sqrt2\, \fft{\del W}{\del \varphi}\,,\qquad
A'=-\fft{1}{\sqrt2\, (D-4)}\, W\,. \label{fo}
\ee

    We lift these equations of motion back to $D$-dimensions,
expressed in terms of the coordinate $\rho$ such that the solution
takes the form
\bea ds_D^2 &=& e^{2u}\, dx^\mu\, dx_\mu + e^{2v}\, \lambda^{-2}\,
d\Omega_2^2 + d\rho^2\,,\nn\\
F_\2&=& \epsilon m\,\lambda^{-2}\, \Omega_\2\,.
\label{metricansatz} \eea
The first-order equations for the functions $u$ and $v$ are given
by
\be \fft{du}{d\rho}=-\fft{2\lambda^2 - \epsilon\,m^2\,
e^{-2v}}{m\, \sqrt{2(D-2)(D-3)}} \,,\qquad
\fft{dv}{d\rho}=-\fft{2\lambda^2 + (D-3)\,\epsilon\, m^2\,
e^{-2v}}{m\,\sqrt{2(D-2)(D-3)}}\,, \ee which have the solution
\be {\rm e}^{2v}={\rm e}^{-\,\frac{4 \lambda^2}{m \sqrt{2
(D-2)(D-3)}}\, \rho} - \frac{(D-3)m^2 \epsilon}{2
\lambda^2}\,,\quad u=-\sqrt{\frac{2(D-2)}{(D-3)^3}}
\frac{\lambda^2}{m}\, \rho-\frac{v}{D-3}\,. \ee
With the new coordinate $r=\lambda^{-1}\, {\rm e}^{v}$, the
solution can be expressed as
\bea
ds_D^2&=&(\lambda\, r)^2
H^{\frac{D-2}{D-3}}dx_{\mu}^2+\frac{dr^2}{(\lambda\, r)^2 H^2}+r^2
d\Omega_2^2\,,\nn\\
F_\2&=&\epsilon\, m\, \lambda^{-2}\Omega_\2 =
\epsilon\, \sqrt{\ft{2(D-1)}{D-3}\, \Lambda}\, \Omega_\2, \label{sabra}
\eea
where $H=1+ \frac{\epsilon\, (D-1)}{\Lambda r^2}$.  It is interesting
to note that, once the cosmological constant is fixed, there is no
free parameter associated with the $F_\2$ field strength.  This is
rather different from the standard brane solution, in which the charge
of a supporting field strength is typically an arbitrary integration
constant. In the present case, it is uniquely determined by a
supersymmetric condition, which translates into the condition
presented below (\ref{W}). For vanishing $\epsilon$, the above
solution is purely gravitational with locally AdS$_D$ geometry.

    The second derivative of the function $u$ with respect to the
co-moving coordinate $\rho$ is given by
\be \fft{d^2u}{d\rho^2} = \fft{2\epsilon\, H}{(D-2)(D-3)\, r^2}\,.
\ee
We find that, for $\epsilon=0$, $d^2u/d\rho^2=0$.  For $\epsilon=-1$,
we have $d^2u/d\rho^2 \le 0$.  In both cases, the solution is regular
everywhere.  On the other hand, if $\epsilon=1$, then $d^2u/d\rho^2
>0$ and the solution is singular.  Thus, there seems to be a clear
correlation between the singularity structure and the sign of
$d^2u/d\rho^2$.  This is rather different from supersymemtric domain
wall solutions where $d^2u/d\rho^2=-(dW/d\phi)^2$ is always
non-positive regardless the singularity structure \cite{naked,
ctheorem}.

    These solutions are among those that have already been found in
\cite{sabra1} by solving the D-dimensional equations of motion
directly \footnote{To recover the notation of \cite{sabra1}, let
$\lambda=\ell$ and $m=\sqrt{\frac{2(D-2)}{D-3}}q\, \ell^2$.}, and the
corresponding Killing spinors have been determined in
\cite{sabra2}. The cases of the $D=5$ magnetic string and the $D=4$
``cosmic monopole'' can be embedded in $N=2$ gauged supergravity
\cite{romans, sabra3, sabra4}.

\section{Single-charge branes in AdS gauged supergravities}

        The above analysis is only applicable in gauged supergravities
in $D=4$ and $D=5$, since in gauged supergravities in $D\ge 6$, the
2-form field strengths always have a dilaton coupling.  In general, a
2-form field strength couples to a dilatonic scalar in the following
fashion:
\be {\cal L}=-\ft14\, e^{-a\, \phi}\, F_\2^2\,, \ee
where the constant $a$ can be parameterized as \cite{stainless}
\be
a^2=\Delta - \fft{2(D-3)}{D-2}\,.\label{avalue}
\ee
Here, $\Delta$ takes the values $4/N$ with $N=1,2,3,4$.  In
\cite{rhamfeld,pbrane}, the $N=1$ solitons are considered as
building blocks, while the $N=2,3,4$ solitons are considered as
threshold binding states of these basic building blocks, since the
field strength associated with $N>1$ can be viewed as linear
combinations of those with $N=1$.  In this section and the next,
we consider magnetic brane solutions in AdS gauged supergravities
supported by a field strength whose dilaton coupling is
characterized as $\Delta=4$ and $2$. Hence, they can be viewed as
single-charge and two-equal-charge branes, respectively.

          The relevant Lagrangian can be expressed as
\be \hat e^{-1}\, \hat {\cal L} = \hat R - \ft12(\del\phi)^2 -
\fft14 e^{-a_1\, \phi}\, \hat F_\2^2 - \hat V(\phi)\,, \ee
where $a_1$ is given by $a$ in (\ref{avalue}).  The scalar
potential can be expressed in terms of a superpotential
\be \hat V=(\fft{\del\hat
W}{\del \phi})^2 - \fft{D-1}{2(D-2)}\,\hat W^2\,,
\ee
and $\hat W$ is given by
\be \hat W=2g\,\Big(\fft{1}{a_1}\, e^{\ft12 a_1\, \phi} -
\fft{1}{a_2}\, e^{\ft12 a_2\, \phi}\Big)\,.
\ee
The value of $a_2$, which can be determined by examining the
gauged supergravities in $D=7,6$ and 5, is given by
\be a_1\, a_2= -\frac{2(D-3)}{D-2}\,.\label{a1a2}
\ee

We can reduce the theory on $d\Omega_2^2$ with the ansatz
\bea
ds_D^2 &=& e^{2\alpha\, \varphi}\, ds_{D-2}^2 +
\lambda^{-2}\, e^{2\beta\,\varphi}\, d\Omega_2^2\,,\nn\\
F_\2&=&\epsilon\, m\, \lambda^{-2}\, \Omega_\2\,,\quad \alpha =
-\fft{1}{\sqrt{(D-2)(D-4)}}\,,\quad \beta = -\ft12 (D-4)\,
\alpha\,. \eea
The $(D-2)$-Lagrangian is given by
\bea
e^{-1}\, {\cal L}&=& R-\ft12{\del\phi}^2 -
\ft12(\del \varphi)^2 -V\,,\nn\\
V&=&\ft12\epsilon^2 m^2\, e^{-a_1\, \phi +
2(D-3)\, \alpha\, \varphi} -
2\epsilon\,\lambda^2\, e^{(D-2)\,\alpha\,\varphi} +
\hat V\, e^{2\, \alpha\,\varphi}\,.
\eea
As in the previous discussion, $\epsilon=1$ for $S^2$ and
$\epsilon=-1$ for $H^2$.  Now it amounts to finding a superpotential
$W$ such that
\be V=(\fft{\del W}{\del \phi})^2 + (\fft{\del W}{\del \varphi})^2
-\fft{D-3}{2(D-4)} \, W^2. \ee
The superpotential exists, provided that $a_1$ and $a_2$ satisfy
(\ref{a1a2}) and that $\lambda$, $g$ and $m$ satisfy
\be
(2\lambda^4 -m^2\, g^2)\, (D-2)\, a_1^2 - 2(D-3)\, m^2\, g^2=0\,.
\label{mglcon}
\ee
The superpotential is given by
\be W= \fft{2a_1\, (D-2)\, \lambda^2\,\epsilon}{g\, (2(D-3) +
a_1^2\, (D-2))}\, e^{-\ft12 a_1\, \phi + (D-3)\,\alpha\, \varphi}
+ e^{\alpha\,\varphi}\,\hat W \,.\label{gensup0} \ee
It is worth noting that the existence of the superpotential only
depends on the condition (\ref{a1a2}), rather than on any specific
value of $a_1$.

      In the case of a single-charge brane, $\Delta=4$ and hence
$a_1^2=\fft{2(D-1)}{D-2}$. The constraint (\ref{mglcon}) becomes
\be
\lambda^2=\sqrt2\, a_1^{-1}\, m\, g\,,
\ee
and the superpotential reduces to
\be W=\ft{1}{\sqrt2}\,\epsilon\,m\,e^{-\ft12 a_1\, \phi +
(D-3)\,\alpha\, \varphi} + \, e^{\alpha\,\varphi}\,\hat W
\,.\label{gensup1} \ee
The first-order equations for the $(D-2)$-dimensional system are
therefore given by
\be \varphi'=\sqrt2\, \fft{\del W}{\del \varphi}\,,\qquad
\phi'=\sqrt2\, \fft{\del W}{\del \phi}\,,\qquad
A'=-\fft{1}{\sqrt2\, (D-4)}\, W\,. \label{genfo} \ee

      Having obtained the first-order equations for the
$D-2$-dimensional domain walls, it is of interest to lift these
equations to $D$ dimensions so that they correspond to magnetic
branes.  The brane solutions have the structure
\bea
ds_D^2 &=&e^{2u}\, dx^\mu\, dx^\nu\, \eta_{\mu\nu} + e^{2v} \,
\lambda^{-2}\,d\Omega_2^2 + d\rho^2\,,\nn\\
F_\2&=&\epsilon\, m\, \lambda^{-2}\,\Omega_\2=
\fft{\epsilon\, a_1}{\sqrt2\, g}\, \Omega_\2\,,\label{braneAdS}
\eea
where the functions $u$ and $v$ and the dilaton $\phi$ depend only on
the coordinate $\rho$.  They satisfy the first-order equations
\bea
\fft{d\phi}{d\rho}&=&\sqrt2 \Big(-\ft1{2\sqrt2}\epsilon\, m\,a_1\,
e^{-\ft12a_1\, \phi - 2v} + \fft{d\hat W}{d\phi}\Big)\,,\nn\\
\fft{dv}{d\rho}&=&-\fft{1}{\sqrt2\, (D-2)}\,\Big(\ft1{\sqrt2}\epsilon\,
m\,(D-3)\,e^{-\ft12a_1\, \phi - 2v} + \hat W\Big)\,,\nn\\
\fft{du}{d\rho}&=&\fft1{\sqrt2\, (D-2)}\, \Big(\ft1{\sqrt2}\epsilon\,
m\,e^{-\ft12a_1\, \phi - 2v}-\hat W\Big)\,.\label{gauged1}
\eea
Note that the $(\phi ,v)$ fields form a closed system.
(\ref{gauged1}) can be solved explicitly by making the coordinate
transformation $d\rho={\rm e}^{-\frac{a_2}{2} \phi}\, dy$.
Defining
\be F\equiv {\rm e}^{\frac{1}{2}(a_2-a_1)\phi}\,,\quad G\equiv
{\rm e}^{\frac{1}{2}(a_1+a_2)\phi +2v}\,, \ee
we find that the first two equations in (\ref{gauged1}) yield
\be G'+\epsilon\, m+\gamma\, G=0\,, \ee
where $\gamma=\frac{4\sqrt{2}\, g}{(D-3)a_1}$. The solution is
\be
G={\rm e}^{-\gamma\, y}-\frac{\epsilon\, m}{\gamma}\,.
\ee
We have absorbed a trivial integration constant by a constant shift in
the coordinate $y$.  The dilaton equation of motion gives
\be \frac{1}{a_1-a_2}\frac{F'}{F}=\frac{\epsilon\, m\,
a_1}{4G}+\frac{g}{\sqrt{2}}(1-F^{-1})\,. \ee
Plugging in $G$ we find that, for $D\ne 5$,
\be F=\frac{{\rm e}^{-\gamma\, y}-\frac{D-3}{D-5} \frac{\epsilon\,
m}{\gamma}+c_1\, {\rm e}^{\ft12(D-5)\,\gamma\, y}}{G}\,.
\ee
This yields
$$
{\rm e}^{\frac{1}{2}(a_1-a_2)\phi}=\frac{{\rm e}^{-\gamma \,
y}-\frac{\epsilon\, m}{\gamma}}{{\rm e}^{-\gamma \, y}- \Big(
\frac{D-3}{D-5} \Big) \frac{\epsilon\, m}{\gamma}+c_1\, {\rm
e}^{\ft12 (D-5)\,\gamma \, y}},
$$
\be
{\rm e}^{2u}=c_2\, {\rm e}^{-\gamma \, y}{\rm
e}^{-\frac{1}{2}(a_1+a_2) \phi}\,,\quad {\rm e}^{2v}=({\rm
e}^{-\gamma \, y}-\frac{\epsilon\, m}{\gamma}){\rm
e}^{-\frac{1}{2} (a_1+a_2) \phi}\,.
\ee
The metric of the solution can be expressed as
\be ds_D^2 = (g\, r)^2\, H^{-\ft{1}{D-2}}\Big(dx_\mu^2 + (1 -
\fft{\epsilon\, m}{\gamma\, (g\, r)^2})\, \lambda^{-2}\,
d\Omega_2^2\Big) + H^{\ft{D-3}{D-2}}\, \fft{4dr^2}{\gamma^2\,
r^2}\,, \label{singlec}
\ee
where
\be
H=\fft{1 - \ft{\epsilon\, m}{\gamma\, (g\, r)^2}}{1 -
\fft{D-3}{D-5}\, \fft{\epsilon\, m}{\gamma\, (g\, r)^2} +
\ft{c_1}{(g\, r)^{D-3}}}\,,\label{genh}
\ee
and the dilaton is given by
\be {\rm e}^{\frac{1}{2}(a_1-a_2)\phi}=H, \ee
with the new coordinate $r=g^{-1}\, e^{-\gamma\, y/2}$. This
solution is valid for $D\ne 5$; we shall present the $D=5$
solution momentarily.  To understand the solution better, it is
instructive to write the function $H=\wtd H/W$, where the
definitions of $\wtd H$ and $W$ can be straightforwardly read off
from (\ref{genh}). Then the metric can be expressed as
\be ds_D^2 = \wtd H^{-\ft{1}{D-2}}\, W^{\ft{1}{D-2}}\,\Big[ (g\,
r)^2\, (dx^\mu\, dx_\mu + \wtd H\, \lambda^{-2}\, d\Omega_2^2) +
\wtd H\, W^{-1}\, \fft{4dr^2}{\gamma^2\, r^2}\Big]\,.
\label{gendmetricsol} \ee
Thus, the solution can be viewed as an intersection of a domain wall,
characterized by the function $H$, and a $(D-3)$-brane, characterized
by the function $\wtd H$.

         Clearly, the asymptotic infinite region of the metric is
$r\rightarrow \infty$, in which case $H\rightarrow 1$ and the metric
behaves as
\be ds_D^2=(g\,r)^2\, (dx^\mu\, dx^\mu + \lambda^{-2}\,
d\Omega_2^2) + \fft{dr^2}{\gamma^2\, r^2}\,.\label{larged1} \ee
If $\epsilon=0$, in which case $d\Omega_2^2$ is the metric of a
2-torus, then the above metric describes locally AdS$_D$
spacetime. For $\epsilon=\pm 1$, the above metric can be viewed as a
domain wall wrapped on $\Omega^2$.

       The full solution with $\epsilon=0$, also obtained in
\cite{gubser}, describes a domain wall.  This can be viewed as a
distribution of branes from the ten or eleven-dimensional point of
view. Hence, this solution corresponds to the Coulomb branch of the
superconformal Yang-Mills theory on the boundary
\cite{gubser,klt,fgpw}.

       For $\epsilon=-1$, the metric is regular everywhere provided
that the constant $c_1=0$.  The corresponding metric interpolates
between AdS$_{D-2}\times H^2$ at small distance to the
AdS$_D$-type metric (\ref{larged1}) at large distance.  When $c_1$
is non-vanishing, the $c_1$ term can dominate at small distance
and hence the metric can become singular.

       For $\epsilon=1$, the solution is always singular at small
distance.  The singularity can be naked if $H$ vanishes at the
singularity but can coincide with the horizon if $H$ is divergent at
the singularity.  The nature of the singularity depends on the
specific value of the constant $c_1$.

        Let us now look at the examples of $D=7, 6, 5$ and $4$ in
detail.

\subsection{3-brane in $D=7$}

   This solution was also obtained in \cite{nunez}.  In this case, the
dilaton and the function $H$ are given by
\be {\rm e}^{\sqrt{\frac{5}{3}}\phi}=H= \fft{1 - \fft{\epsilon\,
m}{\gamma\, (g\, r)^2}}{1 - \fft{2\epsilon\, m}{\gamma\, (g\,r)^2}
+ \fft{c_1}{(g\, r)^4}}\,.\label{d7h} \ee
For $\epsilon=-1$ and $c_1=0$, the metric is regular everywhere, which
approaches AdS$_5\times H^2$ at small distance.  For $c_1\neq 0$, the
metric has a singularity which is naked if $c_1>0$ and which coincides
with a horizon if $c_1 <0$. The structure of the singularity is
qualitatively the same as the domain wall singularity corresponding to
$\epsilon=0$.

     If we set $c_1=0$ and rescale the coordinates as the following:
\be\label{resc}
g\,r \rightarrow k\, g\, r\,,\qquad x^\mu \rightarrow \ft{1}{k}\,
x^\mu\,,
\ee
then by sending $k\rightarrow 0$ we obtain the geometry
AdS$_5\times H^2$ with a constant dilaton.  If we instead take
$c_1=k^2\,\td c_1\,m/\gamma$ then, after sending $k\rightarrow 0$,
we obtain the solution
\bea
ds_7^2 &=& \fft{(g\,r)^2}{ H^{1/5}}\, dx^\mu\, dx_\mu +
\fft{m}{\gamma\,\lambda^2H^{1/5}}\, d\Omega_2^2 +
 H^{4/5}\,\fft{4dr^2}{\gamma^2\,r^2}\,,\nn\\
H^{-1} &=& 2 + \fft{\td c_1}{(g\,r)^2}\,. \label{d7h2}
\eea
When $r\rightarrow \infty$, the metric becomes AdS$_5\times H^2$,
where AdS$_5$ approaches the boundary.  Depending on the sign of ${\td
c}_1$, the metric approaches a singularity at $r=0$ or at some finite
value $r=r_0$.  In both cases, the singularity coincides with a
horizon. However, for ${\td c}_1>0$ the metric components of $H^2$
diverge while, for ${\td c}_1<0$, they vanish.

     This is rather different from our previous solution (\ref{d7h})
with $c_1=0$, in that the structure of AdS$_5\times H^2$ occurs at the
asymptotic region instead of at the horizon.  It is rather interesting
that the AdS$_5\times H^2$ can be in both the IR region, as in
(\ref{d7h}), and in the UV region, as in (\ref{d7h2}). This indicates
that the stationary solution AdS$_5\times H^2$ lies on the inflection
point of the modulus space.

      For $\epsilon=1$, the function $H$ has a simpler form when
$c_1=c_1^*$, where $c_1^*=m^2/\gamma^2$. In this case,
\be
H=\Big(1 - \fft{m}{\gamma\, (g\, r)^2}\Big)^{-1}\,.
\ee
Thus, the singularity at small distance coincides with the
horizon. Clearly, for $c_1\le c_1^*$, the denominator of $H$ in
(\ref{d7h}) approaches zero first and hence the singularity coincides
with the horizon.  On the other hand, for $c_1> c_1^*$ the numerator
of $H$ in (\ref{d7h}) approaches zero first and hence the singularity
is naked.

\subsection{Membrane in $D=6$}

    In this case, the dilaton and the function $H$ are given by
\be\label{H6} {\rm e}^{\sqrt{\frac{8}{5}}\phi}=H= \fft{1 -
\fft{\epsilon\, m}{\gamma\, (g\, r)^2}}{1 - \fft{3\epsilon\,
m}{\gamma\, (g\,r)^2} + \fft{c_1}{(g\, r)^3}}\,.\label{d6h} \ee
The singularity structure is similar to that of $D=7$. For $\e =
-1$ and $c_1=0$, we recover a smooth solution that goes to
AdS$_4\times H^2$ in the IR limit. For $c_1>0$, the denominator of
$H$ in (\ref{H6}) is strictly positive and we have a naked
singularity. On the other hand, for $c_1<0$ the singularity is
located on the horizon.

If we set $c_1=0$ and perform the rescaling (\ref{resc}), then for
$k\ra 0$ we obtain the geometry AdS$_4\times H^2$ with a constant
dilaton. If we instead take $c_1 = k\, \tilde c_1\, m/\gamma$, then
after sending $k\ra 0$ we obtain the solution
\bea
ds_6^2 &=& \fft{(g\,r)^2}{ H^{1/4}}\, dx^\mu\, dx_\mu +
\fft{m}{\gamma\,\lambda^2H^{1/4}}\, d\Omega_2^2 +
 H^{3/4}\,\fft{4dr^2}{\gamma^2\,r^2}\,,\nn\\
H^{-1} &=& 3 + \fft{\td c_1}{g\,r}\,, \label{d6h2} \eea
which approaches AdS$_4\times H^2$ in the asymptotic region, where
AdS$_4$ approaches the boundary. The short-distance behavior is
dependent on ${\tilde c}_1$. More precisely, while the singularity
coincides with the horizon for positive and negative ${\td c}_1$, for
${\td c}_1>0$ the length parameter of $H^2$ diverges at the
singularity. On the other hand, for ${\td c}_1<0$ it vanishes.  Thus,
as in the case of $D=7$, the AdS$_4\times H^2$ solution appears in
both the IR and UV regimes.

Finally, the case of $\epsilon =1$ can be described in terms of the
parameter $c_1^*=2 (m/\gamma)^{3/2}$. For $c_1 > c_1^*$ there is a
naked singularity while, for $c_1 \le c_1^*$, the singularity
coincides with a horizon.

\subsection{String in $D=5$}

One can find the string solution in a manner analogous to the
previous cases, with the coordinate transformation $d\rho={\rm
e}^{\frac{\phi}{\sqrt{6}}}\, dy$. The resulting metric has the
form (\ref{gendmetricsol}) for $D=5$ but with the dilaton and the
function $H$ given by
\be {\rm e}^{\sqrt{\frac{3}{2}}\phi}=H=\fft{1 - \fft{\epsilon\,
m}{\sqrt3\, g^2 r}}{ 1 + (c_1 - \fft{\epsilon\, m}{\sqrt3\, g}\log
(g\, r))\, (g\, r)^{-1}}\,, \label{5H} \ee
where $g\, r={\rm e}^{-\sqrt{3}g\, y}$.  This solution was
obtained and analyzed in \cite{nunez}.  Here, we make a further
observation for $\epsilon=-1$ that if we make the rescaling
(\ref{resc}), accompanied by $c_1=(\td c_1 - \log k) m/(\sqrt3
g)$, then after sending $k\rightarrow 0$ the function $H$ becomes
$H^{-1}=\td c_1 + \log (g\,r)$ and the metric (\ref{singlec})
becomes
\be ds_5^2 = \fft{(g\,r)^2}{H^{1/3}}\, dx^\mu\, dx_\mu +
\fft{m}{\gamma\, \lambda^2\,H^{1/3}}\, d\Omega_2^2 +
H^{2/3}\,\fft{4dr^2}{\gamma^2\,r^2}\,. \ee

\subsection{Black hole in $D=4$}

In this case we have
\be {\rm e}^{\sqrt{\frac{4}{3}}\phi}=H=\fft{1 -\fft{\epsilon\,
m}{\gamma\, (g\,r)^2}}{1 + \fft{c_1}{g\, r} + \fft{\epsilon\,
m}{\gamma\, (g\, r)^2}}\,. \label{d4h} \ee
The metric is clearly singular regardless of the sign of
$\epsilon$. For $\epsilon=-1$, the singularity always coincides
with the horizon. For $\epsilon=1$, the singularity is naked if
$c_1> c_1^*\equiv -2 (m/\gamma)^{1/2}$ but coincides with the
horizon if $c_1 \le c_1^*$.

     For $\epsilon=-1$, we can make the rescaling (\ref{resc})
accompanied by $c_1=m\,\td c_1/(k\, \gamma)$.  Then, after sending
$k\rightarrow 0$, the function $H^{-1}=\td c_1\, g\,r-1$ and the
metric (\ref{singlec}) becomes
\be ds_4^2 = -\fft{(g\,r)^2}{H^{1/2}}\, dt^2 + \fft{m}{\gamma\,
\lambda^2\,H^{1/2}}\, d\Omega_2^2 +
H^{1/2}\,\fft{4dr^2}{\gamma^2\,r^2}\,. \ee
Thus, our solution is more general than the
previously-known AdS black hole \cite{bh5,tenauthors}.

\section{Two-equal-charge branes in AdS gauged supergravities}

           For the case of a brane with two equal charges, $\Delta=2$
and hence $a_1^2=2/(D-2)$. The constraint (\ref{mglcon}) becomes
\be
\lambda^2=a_1^{-1}\, m\,g \,,
\ee
and the superpotential (\ref{gensup0}) becomes
\be W=\epsilon\,m\,e^{-\ft12 a_1\, \phi + (D-3)\,\alpha\, \varphi}
+ \, e^{\alpha\,\varphi}\,\hat W \,.\label{gensup2} \ee

         The first-order equations for the $(D-2)$-dimensional system
are therefore given by (\ref{genfo}). Having obtained the first-order
equations for the domain walls, it is of interest to lift these
equations back to give rise to the first-order equations for the
magnetic branes, whose structure is given by (\ref{braneAdS}).  The
higher-dimensional equations of motion are given by
\bea
\fft{d\phi}{d\rho}&=&\sqrt2 \Big(-\ft12 \epsilon\, m\,a_1\,
e^{-\ft12a_1\, \phi - 2v} + \fft{d\hat W}{d\phi}\Big)\,,\nn\\
\fft{dv}{d\rho}&=&-\fft{1}{\sqrt2\, (D-2)}\,\Big(\epsilon\,
m\,(D-3)\,e^{-\ft12a_1\, \phi - 2v} + \hat W\Big)\,,\nn\\
\fft{du}{d\rho}&=&\fft1{\sqrt2\, (D-2)}\, \Big(\epsilon\,
m\,e^{-\ft12a_1\, \phi - 2v}-\hat W\Big)\,.\label{gauged2}
\eea
Note that the $(\phi ,v)$ fields form a closed system. We can find
exact solutions of this system only in particular cases. In the case
of $D=4$, the system can be solved exactly but the form of the
solution is not very illuminating \cite{gauntlett4}. However, a simple
partial solution for $D=4$ is given by
\be \phi-2v=4g\, y\,,\qquad \ft{d\rho}{dy}={\rm e}^{-\phi/2}. \ee

\subsection{Domain wall solution}

For $\epsilon=0$, and therefore vanishing flux, the first-order
equations (\ref{gauged2}) can be solved straightforwardly by
making the coordinate transformation
\be d\rho=dx\, {\rm e}^{\frac{D-4}{4}a_1\, \phi}\,. \ee
After applying a second coordinate transformation, $(g\,
r)^{\frac{D-3}{2}}={\rm sinh}(-\frac{g}{\sqrt{2}a_1}x)$, the
solution can be expressed as
$$
ds_D^2=(g\, r)^2
H^{\frac{2}{(D-3)(D-2)}}(dx_{\mu}^2+\lambda^{-2}d\Omega_2^2)+
\frac{dr^2}{(g\, r)^2 H^{\frac{2}{D-2}}},
$$
\be {\rm e}^{\phi}=H^{a_1}, \label{epads} \ee
where $H=1+1/(g\, r)^{D-3}$ and $r$ has absorbed a constant
factor. In the above, we have used the convention that $g<0$. This
is a domain wall solution which was obtained in \cite{gubser}.

\subsection{Asymptotic boundary region}

For $\ep=\pm 1$, we do not find analytical solutions.  Here, we
shall present the asymptotic behavior at the boundary of the
metric. Making the coordinate transformation $d\rho=dy\, {\rm
e}^u$, we express the metric (\ref{braneAdS}) as
\be ds_D^2={\rm e}^{2u} (dx^{\mu} dx^{\nu}\eta_{\mu \nu}
+dy^2)+{\rm e}^{2v}\lambda^{-2}d\Omega_2^2. \label{expansion} \ee
When $y\rightarrow 0$ the boundary conditions are $u(y)\sim
v(y)\sim -{\rm log}\, y$, and $\phi \rightarrow 0$. The solution
of (\ref{gauged2}) can be solved near $y=0$ as an expansion. For
$D \ne 5$, the result is
\bea e^{2u}&=&\fft{(D-3)^2}{(D-2)\,g^2\, y^2}\,\Big(1 +
\fft{2\,\epsilon\, m\,g}{3c^2\, (D-3)\,\sqrt{2(D-2)}}\, y^2 +
\cdots\Big)\,,\nn\\
e^{2v}&=& \fft{(D-3)^2\, c^2}{(D-2)\,g^2\,y^2}\,
\Big(1-\fft{\epsilon\, m\,g\,(3D-8)}{3c^2\,(D-3)\,\sqrt{2(D-2)}}\,
y^2 + \cdots\Big)\,,\nn\\
e^\phi &=& 1 + \fft{2\epsilon\,m\,g}{c^2\,(D-3)(D-5)}\, y^2 +
\cdots\,, \eea
where $c$ is an arbitrary integration constant which measures the
relative scale size of $d\Omega_2^2$ and $dx^\mu\, dx_\mu$.  When
$D=5$, the solution of (\ref{gauged2}) is given by the expansion
\bea e^{2u} &=& \fft{4}{3g^2\,y^2}\, (1 +
\fft{\epsilon\,m\,g}{3\sqrt6\,c^2}\,y^2
+\cdots)\,,\nn\\
e^{2v} &=& \fft{4c^2}{3g^2\, y^2}\,
(1 -\fft{7\epsilon\,m\,g}{6\sqrt6\,c^2}\,
y^2 + \cdots)\,,\nn\\
\phi &=& -\ft12 c^{-2}\, \epsilon\,m\,g\, y^2\,\log(y)\,\cdots\,.
\label{twocasym} \eea
In this case, the field $\phi$ is dual to an operator of dimension
$\Delta=2$ \cite{nunez}. The solutions of the two-dimensional wave
equation go as $r^2$ and $r^2\, {\rm log}\, r$. Notice that $\phi$ has
the second solution only for $D=5$, which is the non-normalizable mode
associated with the operator dual to the field $\phi$ turned on.

        Having obtained the boundary solution, it is of interest to
examine how the solution flows into the bulk region.  In the case
of $\epsilon=1$, since there is no AdS$_{D-2}\times S^2$
fixed-point in the system, the solution can only flow into a
singularity in the bulk. For the case of $\epsilon=-1$, there is a
AdS$_{D-2}\times H^2$ fixed-point and it is natural to expect that
the solution flows into this fixed-point in the bulk.  We shall
examine this in the following.

\subsection{AdS$_{D-2}\times H^2$}

For $\epsilon =-1$, there are fixed-point solutions with constant
$v$ and $\phi$ for $D \ge 5$; they are given by
$$
{\rm e}^{-2v}=\frac{2g}{m\, a_1\, (D-4)}\Big(
\frac{D-4}{D-3}\Big)^{a_1^2}\,,\quad {\rm e}^{\phi}=\Big(
\frac{D-4}{D-3}\Big)^{a_1}\,,
$$
\be u=-\frac{\sqrt{2}g}{a_1\, (D-4)}\Big(
\frac{D-4}{D-3}\Big)^{a_1^2/2}\rho\,, \label{Dsolution} \ee
which have the geometry AdS$_{D-2}\times H^2$. For $D=5$, this was
found in \cite{nunez}. For $\epsilon=1$, such a solution would be
complex.

\subsection{Interpolating from AdS$_{D-2}\times H^2$ to AdS$_D$}

       It is of interest to study whether these fixed-point
solutions lie in the IR or UV region.  If AdS$_{D-2}\times H^2$
lies in the IR region, then it can smoothly flow to the UV region
of the AdS$_{D}$-type solution (\ref{twocasym}).  On the other
hand, if AdS$_{D-2}\times H^2$ lies in the UV region, then it must
instead flow into a singularity.

       We study this issue by examining the flow of the solution using
the Taylor expansion method.  If we assume that there exists a
solution which is AdS$_{D-2}\times H^2$ at the asymptotic boundary,
then we can Taylor expand to find the next order in the solution
slightly away from the boundary.  We find the following:
\bea e^{2u} &=& \fft{(D-4)^{\fft{2(D-3)}{D-2}}\,
(D-3)^{\fft2{D-2}}}{
(D-2)\, g^2\, y^2}\,(1 + c\, y^n + \cdots)\,,\nn\\
e^{2v} &=& \fft{m\, (D-4)^{\fft{D-4}{D-2}}\, (D-3)^{\fft2{D-2}}}{
\sqrt{2(D-2)}\, g}\, (1 - \ft12 c\,(D-4)\, y^n + \cdots)\,,\nn\\
e^{\fft{\phi}{\sqrt{2(D-2)}}} &=&
\Big(\fft{D-4}{D-3}\Big)^{\fft1{D-2}}\, \Big(
1\nn\\
&& + \ft18c\, (D^2 - 3D -2 + (D-2)\,\sqrt{D^2 - 2D + 7})\, y^n +
\cdots\Big)\,,\label{h2singular}
\eea
where $n=\ft12 (D-5 + \sqrt{D^2 - 2D - 7})$ and the coordinate $y$
is defined by $d\rho=e^{u}\, dy$.  The constant $c$ is a free
parameter determining how fast the solution flows away from the
AdS$_{D-2}$ boundary.  The solution will flow into a singularity
in the IR region.

\begin{figure}
   \epsfxsize=4.0in \centerline{\epsffile{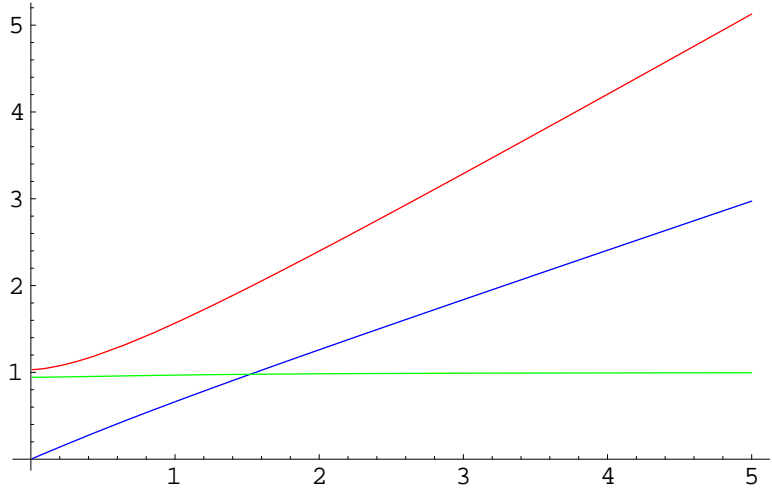}}
   \caption[FIG. \arabic{figure}.]{$e^u$ (blue), $e^{v}$ (red) and
   $\phi$ (green) for a smooth solution that runs from AdS$_5\times
   H^2$ at the horizon to an AdS$_7$-type geometry in the asymptotic
   region.}  \label{2cgenfig}
\end{figure}

      Now, we will instead assume that there exists a solution with
AdS$_{D-2}\times H^2$ on its horizon. Then, expanding away from the
horizon yields
\bea
e^{2u}&=& (D-4)^{-\fft{2(D-3)}{D-2}}\, (D-3)^{\fft{2}{D-2}}\,
(D-2)\, g^2\, r^2\, (1 + c\, r^n + \cdots)\,,\nn\\
e^{2v} &=& \fft{m\, (D-4)^{\fft{D-4}{D-2}}\, (D-3)^{\fft2{D-2}}}{
\sqrt{2(D-2)}\, g}\, (1 - \ft12 c\,(2D-7 + \sqrt{D^2-2D-7})\, r^n
+ \cdots)\,,\nn\\
e^{\fft{\phi}{\sqrt{2(D-2)}}} &=&
\Big(\fft{D-4}{D-3}\Big)^{\fft1{D-2}}\, \Big(1\nn\\
&& + \ft18c\, (D^2 - 5D + (D-4)\,\sqrt{D^2 - 2D + 7})\, r^n +
\cdots\Big)\,,\label{h2regular}
\eea
where $n=\ft12(5-D+\sqrt{D^2-2D-7})$ and the coordinate $r$ is
defined by $d\rho = e^{-u}\, dr$.  The constant $c$ is again a
free parameter determining how fast the solution flows away from
the AdS$_{D-2}$ horizon.  In fact, the solution smoothly flows to
the AdS$_{D}$-type boundary solution (\ref{twocasym}).  We verify
this numerically by using (\ref{h2regular}) as our initial data.
For example, Fig.~\ref{2cgenfig} shows the functions $e^u$, $e^v$
and $e^{\phi/\sqrt{10}}$ for $D=7$.  We see that the dilaton,
although not constant, is finite. The functions $e^u$ and $e^v$
become linearly dependent on the coordinate when the coordinate
increases as governed by (\ref{twocasym}) with $y\rightarrow 1/y$.

     To conclude, we find that for two-equal-charge branes
there are two types of solutions.  The first type has
AdS$_{D-2}\times H^2$ in the UV boundary region but is singular in
the IR region.  The second type interpolates between the
AdS$_D$-type solution in the UV region (\ref{twocasym}) and the
AdS$_{D-2}\times H^2$ solution in the IR region.

\section{Two-charge 3-brane in $D=7$}

     The maximal gauged supergravity in $D=7$ has $SO(5)$ gauge
fields.  We consider the truncation to the two diagonal $U(1)$
subsector.  The relevant $D=7$ Lagrangian is given by
\be
\hat e^{-1}{\cal L}_7 = \hat R - \ft12 (\del \phi_1)^2 -
\ft12 (\del \phi_2)^2 -
\hat V -\ft14 \sum_{i=1}^2 X_i^{-2}\, (\hat F_\2^i)^2\,,
\ee
where $X_i=e^{\ft12\vec a_i\cdot \vec \phi}$ with
\be
\vec a_1=(\sqrt2\,, \sqrt{\ft25})\,,\qquad
\vec a_2=(-\sqrt2\,, \sqrt{\ft25})\,.
\ee
The scalar potential $\hat V$ is given by \cite{tenauthors}
\be \hat V =g^2\, (-4X_1\, X_2 - 2 X_0\, X_1 - 2 X_0\, X_2 + \ft12
X_0^2) \,, \ee
where $X_0=(X_1\, X_2)^{-2}$.  The potential can be expressed in
terms of a superpotential given by
\be
\hat W=\fft{g}{\sqrt2}\, (X_0 + 2X_1 + 2 X_2)\,.
\ee

    We can consistently reduce the theory on the $S^2$ with the same
metric ansatz as before but specializing to $D=7$. The ansatz for
the two $U(1)$ 2-form field strengths is given by
\be F_\2^i= \epsilon\, m_i\, \lambda^{-2}\, \Omega_\2\equiv
\ft12\epsilon\,q_i\, g^{-1}\, \Omega_\2\,. \ee
We obtain the $D=5$ scalar potential
\be
V=\ft12\epsilon^2 \,(\sum_{i=1}^2 m_i^2\, X_i^{-2})\,
e^{8 \alpha\, \varphi} -
2 \epsilon \lambda^2\, e^{5\alpha\, \varphi} + \hat V\,
e^{2\alpha\, \varphi}\,.
\ee
It is straightforward to obtain the corresponding superpotential,
which is given by
\be
W=\fft{\epsilon}{\sqrt2} (\sum_{i=1}^2 m_i\, X_i^{-1} )\,
e^{4\alpha\, \varphi} + \hat W\, e^{\alpha\, \varphi}\,,
\ee
provided that the constraint
\be
\lambda^2 = (m_1 + m_2)\,g\label{2cd7con}
\ee
is satisfied.  Thus, the charge parameters $q_i$ have the
constraint $q_1 + q_2=2$.  The first-order equations of the
five-dimensional domain wall are given by
\be
\phi_i'=\sqrt2\, \fft{\del W}{\del \phi_i}\,,\qquad
\varphi'=\sqrt2\, \fft{\del W}{\del \varphi}\,,\qquad
A'=-\fft{1}{3\sqrt2}\, W\,.\label{d7to5fo}
\ee
Lifting the equations back to $D=7$ yields the equations of motion
for the 3-brane:
\bea
\fft{d\vec \phi}{d\rho} &=& \sqrt2\,\Big(-\fft{\epsilon}{2\sqrt2}\,
(m_1\, \vec a_1\, X_1^{-1} + m_2\, \vec a_2\, X_2^{-1})\,
e^{-2v} + \fft{d\hat W}{d\vec \phi}\Big)\,,\nn\\
\fft{dv}{d\rho} &=& -\fft{1}{5\sqrt2}\, \Big(2\sqrt2\,
\epsilon\,(m_1\, X_1^{-1} + m_2\, X_2^{-1})\, e^{-2v} +
\hat W\Big)\,,\nn\\
\fft{du}{d\rho} &=& \fft{1}{5\sqrt2}\, \Big(
\fft{\epsilon}{\sqrt2}\,(m_1\, X_1^{-1} + m_2\, X_2^{-1})\,
e^{-2v} - \hat W\Big)\,. \label{7} \eea

        In the case of $m_2=0$, it is consistent to set
$\phi_1=\sqrt{5} \phi_2$, and the system (\ref{7}) reduces to the
$D=7$ single-charge system discussed in section 3.1. After taking
$g\rightarrow \sqrt{5/6}\, g$ and $\phi_2 \rightarrow \phi
/\sqrt{6}$, the exact solution is given by (\ref{gendmetricsol})
and (\ref{d7h}). This solution was obtained in \cite{nunez}.  For
$m_1=m_2$, the system reduces to the $D=7$ two-equal-charge system
discussed in section 3.2.  For general $m_i$, we do not find
analytical solutions and so it is instructive to study the
solution using numerical approach.

\subsection{General asymptotic region}

    We begin by determining the asymptotic behavior in the AdS$_7$-type
boundary. As discussed in the previous section, we can express the
metric in the form (\ref{expansion}) by making the coordinate
transformation $d\rho=dy\, {\rm e}^u$. We have the boundary conditions
$u(y)\sim v(y)\sim -{\rm log}\, y$, and $\phi \rightarrow 0$ for small
$y$. The leading terms in the Taylor expansion of the solution of
(\ref{7}) are given by
\bea e^{2u} &=& \fft{4}{g^2\, y^2}\, ( 1 +
\fft{\epsilon\, (m_1 + m_2)\, g}{30 c^2}\, y^2 + \cdots)\,,\nn\\
e^{2v} &=& \fft{4c^2}{g^2\, y^2}\, ( 1 -
\fft{13\epsilon\, (m_1+m_2)\, g}{60c^2}\, y^2 + \cdots)\,,\nn\\
e^{\ft{\phi_1}{\sqrt2}} &=& 1 + \fft{\epsilon\, (m_1 -
m_2)\,g}{8c^2}\, y^2 + \cdots\,,\\ e^{\fft{\phi_2}{\sqrt{10}}} &=&
1 + \fft{\epsilon\, (m_1 + m_2)\,g}{40c^2}\, y^2 + \cdots\,,\nn
\label{2cd7asym} \eea
where $c$ is an arbitrary integration constant.  The normalizable
solutions for $\phi_1$ and $\phi_2$ imply that the dual operators are
not turned on.  Next, we examine how these solutions flow from the
above boundary into the bulk.

\subsection{AdS$_5\times H^2$ and AdS$_5\times S^2$}

For $\epsilon =\pm 1$, solutions to the equations (\ref{7}) with
constant $v$ and ${\vec \phi}$ are given by
\bea
{\rm e}^{\sqrt{2}\phi_1}&=&\frac{m_2-m_1 \pm \sqrt{(m_2-m_1)^2+m_1\,
m_2 }}{m_2}\,,\quad {\rm
e}^{-\sqrt{\frac{5}{2}}\phi_2}=\frac{4}{3} {\rm
cosh}(\phi_1/\sqrt{2})\,,\nn\\
{\rm e}^{-2v}&=& -\frac{g\, {\rm
e}^{-\frac{3}{\sqrt{10}}\phi_2}}{\epsilon\, (m_1\, {\rm
e}^{-\frac{\phi_1}{\sqrt{2}}}+m_2\, {\rm
e}^{\frac{\phi_1}{\sqrt{2}}})}\,,\quad u=-\frac{g}{2}{\rm
e}^{-\frac{4}{\sqrt{10}}\phi_2}\rho\equiv
-\fft{2}{R_{\rm ads}}\,\rho\,. \label{ads5s2} \eea
This solution is discussed in detail in \cite{adsletter}.  Note that
the AdS radius $R_{\rm ads}$ depends on two variables, namely $g$ and
the ratial of the charge parameter $m_1/m_2$.  This is because the
charges of this system are constrained by $q_1 + q_2=2$, as discussed
earlier.  It is invariant under the simultaneous interchanges
$m_1\leftrightarrow m_2$ and $\phi_1\leftrightarrow -\phi_1$.  The
reality conditions of the solution constrain the constants $m_i$ and
$g$, as well as the choice of $\pm$ in the solution.  Let us first
consider the case $\ep=-1$, corresponding to $d\Omega_2^2$ as the
metric of a unit (non-compact) hyperbolic 2-plane.  In this case, the
reality of the solution implies that $m_1\, m_2 \ge 0$. This includes
the choices of $m_1=0$ (or $m_2=0$) and $m_1=m_2$, which were
discussed in \cite{nunez}.  The first case gives rise to ${\cal N}=4$
supersymmetry in $D=5$, while the second case gives rise to ${\cal
N}=2$ supersymmetry.

    Now let us look at the choice of $\epsilon=+1$, corresponding to
$d\Omega_2^2$ as the metric of $S^2$.  In this case, the reality
conditions for (\ref{ads5s2}) imply that $m_1\, m_2<0$.  The
condition (\ref{2cd7con}) implies further that $m_1\ne -m_2$.
Therefore, the AdS$_5\times S^2$ solution can only have ${\cal
N}=2$ supersymmetry but cannot arise from the pure $D=7$ minimal
gauged supergravity.

        If we define a charge parameter $q=2m_1/(m_1 + m_2)$, then the
condition for having $S^2$ versus $H^2$ can be summarized as the
following:
\bea
q\in [0,2] &\Longrightarrow& H^2\,,\nn\\
q\in (-\infty,0)\,\, \hbox{or}\,\, (2,\infty) &\Longrightarrow& S^2\,.
\label{s2h2rule}
\eea
The solution was lifted to $D=11$ in \cite{adsletter}, and the
corresponding metric describes a warped product of AdS$_5$ with a
six-dimensional space which can be viewed as $S^4$ bundle over
$H^2$ or $S^2$.  In the latter case, the solution provides a
concrete example of a supersymmetric and smooth compactification
of M-theory to AdS$_5$ \cite{adsletter}.

\subsection{Interpolating from AdS$_5 \times H^2$ to AdS$_7$}

      As discussed above, AdS$_5\times H^2$ arises for $m_1\,
m_2\ge 0$ and $\epsilon=-1$.  The special cases of a single
non-vanishing $m_i$ or $m_1=m_2$ have been discussed in sections 3
and 4, respectively . We now consider the general case of $m_i$
with $m_1\, m_2>0$. As in the previous cases, AdS$_5\times H^2$
can occur at either the boundary or the horizon of the AdS$_5$. In
order to demonstrate this, we use a special case of $m_1=16$,
$m_2=6$ and a positive sign in (\ref{ads5s2}). For simplicity, we
also set $g=1$. As a boundary solution, the Taylor expansion of
the metric components and the scalars goes as
\bea
e^{2u}&=&\ft{18}{5}\, (\ft{18}{125})^{1/5}\, y^{-2}\, (1 +
c\, y^n + \cdots)\,,\nn\\
e^{2v} &=& 6 (450)^{1/5}\, (1 -\frac{3}{2}c\, y^n + \cdots)
\,,\nn\\
e^{\ft{\phi_1}{\sqrt2}} &=& \sqrt{\ft23}\, (1 -
\fft{75c\,(n+1)}{2(25n-68)}\,
y^n + \cdots)\,,\nn\\
e^{\ft{\phi_2}{\sqrt{10}}} &=& (\ft{27}{50})^{1/10}\,
(1 -\fft{c\,(n+1)}{2(n-4)}\,y^n + \cdots)\,,
\eea
where the coordinate $y$ is defined to be $d\rho=e^{u}\, dy$ and $c$
is an arbitrary integration constant.  The constant $n$ can take two
values: $n=2$ or $n=1 + \ft15\sqrt{193}$.  Clearly, this solution will
flow into a singularity in the bulk.

\begin{figure}
   \epsfxsize=4.0in \centerline{\epsffile{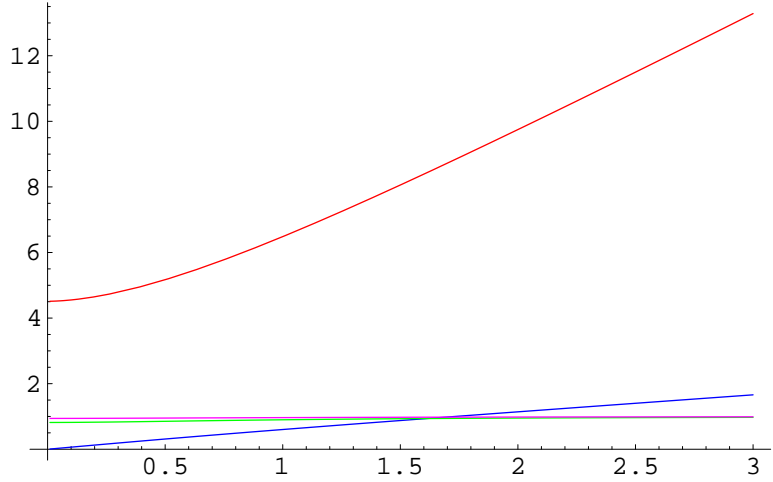}}
   \caption[FIG. \arabic{figure}.]{Plots of $e^u$ (blue), $e^{v}$
   (red), $e^{\phi_1/\sqrt2}$ (green) and $e^{\phi_2/\sqrt{10}}$
   (purple) in a smooth solution that runs from AdS$_\5\times H^2$ at
   the horizon to the AdS$_7$-type geometry in the asymptotic region.
   Note that the two scalar curves are almost identical on this scale.
   $m_1=16, m_2=6$, $g=1$, and $c=-0.1$.} \label{2cd7h2fig}
\end{figure}

     On the other hand, as a solution near the horizon, we find that
the Taylor expansion of the next to leading order is given by
\bea
e^{2u}&=&\ft{18}{5}\, (\ft{18}{125})^{1/5}\, r^2\, (1 +
c\, r^n + \cdots)\,,\nn\\
e^{2v} &=& 6 (450)^{1/5}\, (1 - \fft{3(2895 + 209\sqrt{193})\,c}{2
(805 + 59\sqrt{193})}\, r^n + \cdots)
\,,\nn\\
e^{\ft{\phi_1}{\sqrt2}} &=&
\sqrt{\ft23}\, (1 - \fft{15\sqrt{193}\,
c}{2(43 + 5\sqrt{193})}\, r^n + \cdots)\,,\nn\\
e^{\ft{\phi_2}{\sqrt{10}}} &=& (\ft{27}{50})^{1/10}\, (1
-\fft{c\,\sqrt{193}}{2(15 + \sqrt{193})}\,r^n + \cdots)\,,
\eea
where $n=\ft15\sqrt{193}-1$ and $c$ is an arbitrary integration
constant.  The coordinate $r$ is defined by $d\rho = e^{-u}\, dr$.  We
can use the above as the initial data to obtain a numerical
solution. Plots of the functions $e^{u}, e^{v}, e^{\phi_1/\sqrt2}$ and
$e^{\phi_2/\sqrt{10}}$ are presented in Fig.~\ref{2cd7h2fig}, which
clearly shows that the solution runs smoothly from AdS$_5\times H^2$
at $r\rightarrow 0$ to the AdS$_7$-type solution (\ref{2cd7asym}) at
$r\rightarrow \infty$.

\subsection{Interpolating from AdS$_5 \times S^2$ to AdS$_7$}

      The AdS$_5\times S^2$ fixed-point arises for $m_1\, m_2 <0$ and
$\epsilon=+1$.  The Taylor expansion analysis indicates that the
solution can occur either in the boundary or the horizon of the
AdS$_5$.  For a concrete example, we choose $m_1=5$, $m_2=-3$, $g=1$
and a negative sign in (\ref{ads5s2}).  The boundary solution behaves
as
\bea e^{2u} &=& (\ft{25}{8})^{2/5}\, y^{-2}\, (1 + c\, y^n +
\cdots) \,,\qquad
e^{2v} = (\ft{625}{2})^{1/5}\, (1 -\ft32c\, y^n + \cdots)\,,\nn\\
e^{\ft{\phi_1}{\sqrt2}} &=& \sqrt5\, (1 + \fft{5(n+1)}{n-10} c\,
y^n + \cdots)\,,\nn\\
e^{\ft{\phi_2}{\sqrt{10}}} &=& (\ft{5}{16})^{1/10}\, (1 -
\fft{n+1}{2(n-4)} c\, y^n + \cdots)\,, \eea
where $n=2$ or $n=1+\sqrt{15}$.  The constant $c$ is an arbitrary
integration constant.  Clearly, the flow of this solution from the
boundary leads to a singularity in the bulk.

\begin{figure}
   \epsfxsize=4.0in \centerline{\epsffile{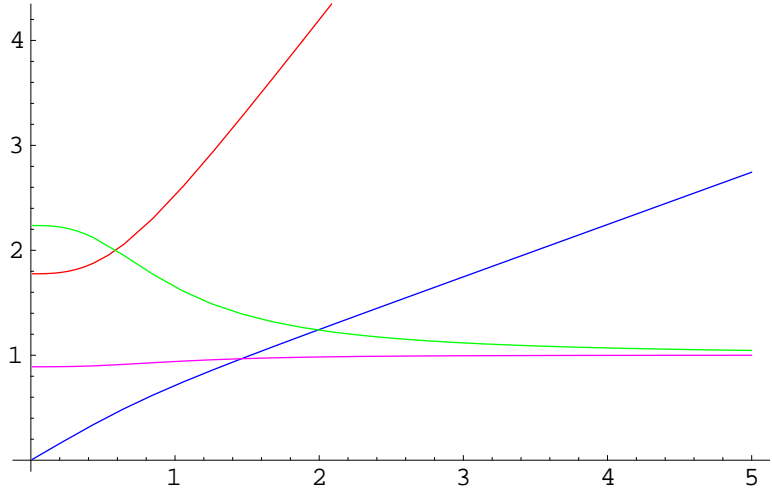}}
   \caption[FIG. \arabic{figure}.]{$e^u$ (blue), $e^{v}$ (red),
   $e^{\phi_1/\sqrt2}$ (green) and $e^{\phi_2/\sqrt{10}}$ (purple) for
   a smooth solution that runs from AdS$_\5\times S^2$ at horizon to
   the AdS$_7$-type geometry in the asymptotic region.  $m_1=5,
   m_2=-3$, $g=1$, and $c=-0.2$.} \label{2cd7s2fig}
\end{figure}

       As a horizon, AdS$_5\times S^2$ has the behavior
\bea
e^{2u} &=& (\ft{64}{625})^{1/5}\, r^2\, (1 + c\, r^n + \cdots)\,,\nn\\
e^{2v} &=& (\ft{625}{2})^{1/5}\, (1 -\fft{3(15 +
4\sqrt{15})\,c}{2(7 +
2\sqrt{15})}\, r^n + \cdots)\,,\nn\\
e^{\ft{\phi_1}{\sqrt2}} &=& \sqrt5\, (1 + \fft{5\sqrt{15}\,c}{9 +
\sqrt{15}}\, r^n + \cdots)\,,\nn\\
e^{\ft{\phi_2}{\sqrt{10}}} &=& (\ft{5}{16})^{1/10}\, (1 -
\fft{\sqrt{15}\, c}{2(3 + \sqrt{15})}\, r^n + \cdots)\,,
\eea
where $n=-1 + \sqrt{15}$, $c$ is an integration constant and $r$
is defined by $d\rho=e^{-u}\, dr$.  We can use this solution as
our initial data for numerical analysis.  The result is plotted in
Fig.~\ref{2cd7s2fig}.

The plot clearly demonstrates that there is a smooth solution
which interpolates between AdS$\times S^2$ at $r\rightarrow 0$ and
the AdS$_7$-type solution (\ref{2cd7asym}) at $r\rightarrow
\infty$. This solution provides a concrete framework with which to
study the M-theory dual to a certain six-dimensional theory whose
IR dynamics is given by a $D=4$, ${\cal N}=1$ superconformal field
theory.

\section{Two-charge membrane in $D=6$}

          The scalar potential in gauged supergravity with two $U(1)$
isometries was obtained in \cite{gubser}.  From this, we deduce
that the relevant Lagrangian involving the two $U(1)$ vector
fields is given by
\be
\hat e^{-1}{\cal L}_6 = \hat R - \ft12 (\del \phi_1)^2 -
\ft12 (\del \phi_2)^2 -
\hat V -\ft14 \sum_{i=1}^2 X_i^{-2}\, (\hat F_\2^i)^2\,,
\ee
where $X_i=e^{\ft12\vec a_i\cdot \vec \phi}$ with
\be
\vec a_1=(\sqrt2\,, \fft1{\sqrt2})\,,\qquad
\vec a_2=(-\sqrt2\,, \fft1{\sqrt2})\,.
\ee
The scalar potential is given by
\be
\hat V=\ft49 g^2\, (X_0^2 - 9 X_1\, X_2 - 6 X_0\, X_1 - 6 X_0\, X_2)\,,
\ee
where $X_0=(X_1\,X_2)^{-3/2}$.  As in the previous case, the scalar
potential can be expressed in terms of a superpotential $\hat W$,
given by
\be
\hat W=\fft{g}{\sqrt2}\,(\ft43 X_0 + 2 X_1 + 2 X_2)\,.
\ee

       We consistently reduce the theory on the $S^2$ or $H^2$ with
the same metric ansatz as before but specializing to $D=6$. The
ansatz for the two $U(1)$ field strengths is given by $F_\2^i=
\epsilon m_i\, \lambda^{-2}\, \Omega_\2\equiv \ft12\epsilon\,q_i\,
g^{-1}\, \Omega_\2$. We obtain the $D=4$ scalar potential
\be
V=\ft12\epsilon^2 \,(\sum_{i=1}^2 m_i^2\, X_i^{-2})\,
e^{6 \alpha\, \varphi} -
2 \epsilon  \lambda^2\, e^{4\alpha\, \varphi} +
\hat V\, e^{2\alpha\, \varphi}\,.
\ee
It is straightforward to obtain the corresponding superpotential,
given by
\be
W=\fft{\epsilon}{\sqrt2} (\sum_{i=1}^2 m_i\, X_i^{-1} )\,
e^{3\alpha\, \varphi} + \hat W\, e^{\alpha\, \varphi}\,.
\ee
As in the $D=7$ case, the constraint $\lambda^2=(m_1 + m_2)\, g$ must
be satisfied, implying $q_1+q_2=2$.  The first-order equations of the
four-dimensional domain-wall has the same form as (\ref{d7to5fo}), but
with $A'=-W/(2\sqrt2)$.  Lifting the equations back to $D=6$ yields
\bea
\fft{d\vec \phi}{d\rho} &=& \sqrt2\,\Big(-\fft{\epsilon}{2\sqrt2}\,
(m_1\, \vec a_1\, X_1^{-1} + m_2\, \vec a_2\, X_2^{-1})\,
e^{-2v} + \fft{d\hat W}{d\vec \phi}\Big)\,,\nn\\
\fft{dv}{d\rho} &=& -\fft{1}{4\sqrt2}\, \Big(\fft3{\sqrt2}\,
\epsilon\,(m_1\, X_1^{-1} + m_2\, X_2^{-1})\, e^{-2v} +
\hat W\Big)\,,\nn\\
\fft{du}{d\rho} &=& \fft{1}{4\sqrt2}\, \Big(
\fft{\epsilon}{\sqrt2}\,(m_1\, X_1^{-1} + m_2\, X_2^{-1})\,
e^{-2v} - \hat W\Big)\,. \label{6} \eea

     In the case of $m_2=0$, it is consistent to set $\phi_1=2
\phi_2$, and the system reduces to the $D=6$ system discussed in
section 3.1. After taking $g\rightarrow \sqrt{4/5}\, g$ and
$\phi_2 \rightarrow \phi /\sqrt{5}$, the exact solution is given
by (\ref{gendmetricsol}) and (\ref{d6h}). For general $m_i$, we do
not find analytical solutions and so it is instructive to study
the system using the numerical approach.

\subsection{Asymptotic boundary region}

We can express the metric in the form (\ref{expansion}) by making
the coordinate transformation $d\rho=dy\, {\rm e}^u$. We have the
boundary conditions $u(y)\sim v(y)\sim -{\rm log}\, y$ and $\phi
\rightarrow 0$ for small $y$. The leading terms in the Taylor
expansion of the solution of (\ref{6}) are given by
\bea e^{2u} &=& \fft{9}{4g^2\, y^2}\, ( 1 +
\fft{\epsilon\, (m_1 + m_2)\, g}{18 c^2}\, y^2 + \cdots)\,,\nn\\
e^{2v} &=& \fft{9c^2}{4g^2\, y^2}\, ( 1 -
\fft{5\epsilon\, (m_1+m_2)\, g}{18c^2}\, y^2 + \cdots)\,,\nn\\
e^{\ft{\phi_1}{\sqrt2}} &=& 1 + \fft{\epsilon\, (m_1 -
m_2)\,g}{3c^2}\, y^2 + \cdots\,,\quad e^{\fft{\phi_2}{\sqrt{8}}} =
1 + \fft{\epsilon\, (m_1 + m_2)\,g}{12c^2}\, y^2 +
\cdots\,,\label{2cd6asym} \eea
where $c$ is an arbitrary integration constant. As in the case of
$D=7$, the normalizable solutions for $\phi_1$ and $\phi_2$ imply that
the dual operators are not turned on. Next, we examine how these
solutions flow from the above boundary into the bulk.

\subsection{AdS$_4\times H^2$ and AdS$_4\times S^2$}

For $\epsilon= \pm 1$, solutions to the equations (\ref{6}) with
constant $v$ and ${\vec \phi}$ are given by
\bea
{\rm e}^{\sqrt{2}\phi_1}&=&\frac{3}{2}\, \frac{m_2-m_1 \pm
\sqrt{(m_2-m_1)^2+\frac{4}{9}m_1\, m_2 }}{m_2}\,,\quad {\rm
e}^{-\sqrt{2}\phi_2}=\frac{3}{2} {\rm
cosh}(\frac{\phi_1}{\sqrt{2}})\,,\nn\\
{\rm e}^{-2v} &=& -\frac{4g\, {\rm
e}^{-\frac{\phi_2}{\sqrt{2}}}}{3\epsilon\, (m_1\, {\rm
e}^{-\frac{\phi_1}{\sqrt{2}}}+m_2\, {\rm
e}^{\frac{\phi_1}{\sqrt{2}}})}\,,\quad u=-\frac{2}{3}g\, {\rm
e}^{-\frac{3}{\sqrt{8}}\phi_2}\rho \equiv
-\fft{2}{R_{\rm ads}}\, \rho, \label{ads4omega2}
\eea
which have the geometry AdS$_4\times H^2$ or AdS$_4 \times S^2$
for $\epsilon=-1$ or $+1$, respectively. These solutions were
discussed in detail in \cite{adsletter} and are similar to one
found in \cite{nunez2}. As in the $D=7$ result, we can define a
charge parameter $q=\fft{2m_1}{m_1 + m_2}$.  We have $H^2$ or
$S^2$ depending on the following conditions:
\bea
q\in [0,2] &\Longrightarrow& H^2\,,\nn\\
q\in (-\infty,0)\,\, \hbox{or}\,\, (2,\infty) &\Longrightarrow&
S^2\,. \label{d6rule} \eea
When $q=0$ or $q=2$, the system has ${\cal N}=4$ supersymmetry.
Otherwise, it has ${\cal N}=2$ supersymmetry.  Note that the AdS$_4$
radius $R_{\rm ads}$ depends on only $g$ and $m_1/m_2$.

          These solutions have been lifted back to $D=10$ massive
supergravity in \cite{adsletter}.  The $D=10$ metric describes a
warped product of Ads$_4$ with an internal metric composed of an
$S^4$ bundle over $S^2$ or $H^2$.

\subsection{Interpolating from AdS$_4 \times H^2$ to AdS$_6$}

      As we discussed above, AdS$_4\times H^2$ arises for $m_1\,
m_2\ge 0$ and $\epsilon=-1$.  The special cases of one
non-vanishing $m_i$ and $m_1=m_2$ have been discussed in sections
3 and 4, respectively . We will now consider the general case of
$m_i$ with $m_1\, m_2>0$. As in the previous cases, AdS$_4\times
H^2$ can occur as an asymptotic geometry in one solution and a
horizon geometry in another.  In order to demonstrate this
explicitly, we take $m_1=5$, $m_2=2$ and a positive sign in
(\ref{ads4omega2}). For simplicity, we also set $g=1$.  As an
asymptotic boundary solution, the Taylor expansions of the metric
components and the scalars go as
\bea e^{2u}&=& (\ft{128}{81})^{1/4}\, y^{-2}\, (1 +
c\, y^n + \cdots)\,,\nn\\
e^{2v} &=& (648)^{1/4}\, (1 -c\, y^n + \cdots)
\,,\nn\\
e^{\ft{\phi_1}{\sqrt2}} &=& \ft{1}{\sqrt{2}}\, (1 -
\ft{12(n+1)}{9n-17}c\,
y^n + \cdots)\,,\nn\\
e^{\ft{\phi_2}{\sqrt{8}}} &=& (\ft{32}{81})^{1/8}\, (1
-\ft{(3n+5)(6n-7)}{2(9n-17)}\,c\, y^n + \cdots)\,, \eea
where the coordinate $y$ is defined to be $d\rho=e^{u}\, dy$ and
$c$ is an arbitrary integration constant.  The constant $n$ can
take two values: $n=1$ or $n=\ft16 (3+\sqrt{185})$.  Clearly, this
solution will flow into a singularity in the bulk.

\begin{figure}
   \epsfxsize=4.0in \centerline{\epsffile{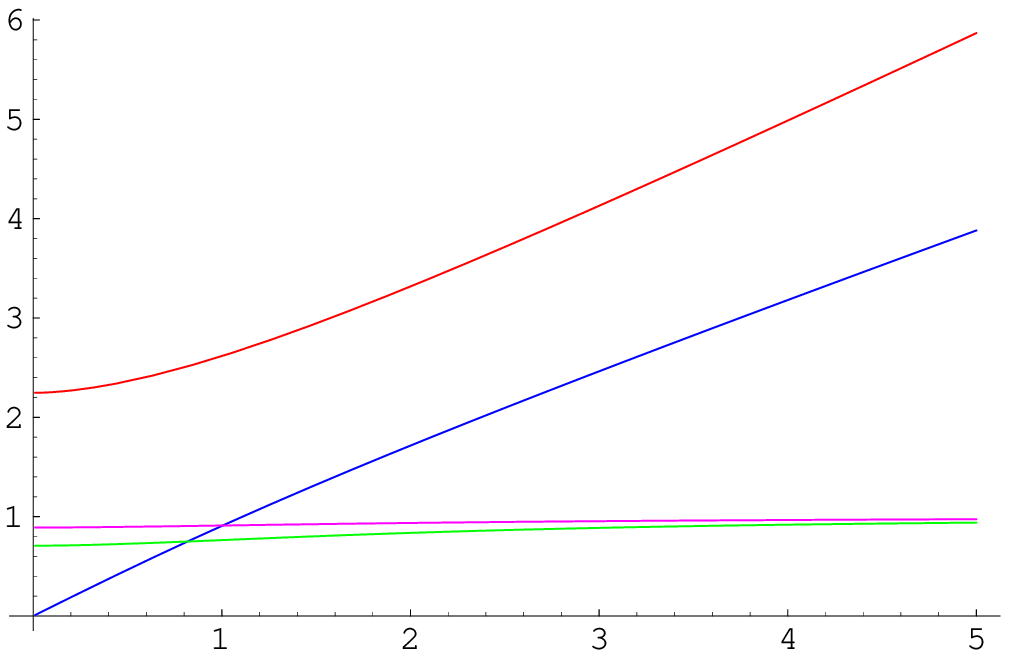}}
   \caption[FIG. \arabic{figure}.]{$e^u$ (blue), $e^{v}$
   (red), $e^{\phi_1/\sqrt2}$ (green) and $e^{\phi_2/\sqrt{8}}$
   (purple) for a smooth solution running from AdS$_4\times
   H^2$ at the horizon to an asymptotic AdS$_7$-type geometry.
   $m_1=5, m_2=-3$, $g=1$, and $c=-0.05$.} \label{2cd6h2fig}
\end{figure}

     On the other hand, for the solution with AdS$_4\times H^2$ at
the horizon, we find the Taylor expansion of the next leading
order to be given by
\bea
e^{2u} &=& (\ft{81}{128})^{1/4}\, r^2\, (1 + c\, r^2 + \cdots)\,,\nn\\
e^{2v} &=& (648)^{1/4}\, (1 - \ft{25n-69}{7n-19}\, c\, r^2 + \cdots)\,,
\nn\\
e^{\ft{\phi_1}{\sqrt2}} &=&\ft1{\sqrt2}\, (1 - \ft{12(n+1)}{9n-17}\,
c\,r^2 + \cdots)\,,\nn\\
e^{\ft{\phi_2}{\sqrt8}} &=& (\ft{32}{81})^{1/8}\, (1 -
\ft{(n+1)\, c}{2(n+3)}\,c\, r^2 + \cdots)\,,
\eea
where $n=\ft16(-3 + \sqrt{185})$ and $c$ is an integration
constant. The coordinate $r$ is defined by $d\rho=e^{-u}\, dr$.
This solution runs smoothly to large distance, where it asymptotes
to the AdS$_6$-type solution (\ref{2cd6asym}).  This can be
demonstrated with numerical calculation, using the above Taylor
expansions as the initial data. In Fig.~\ref{2cd6h2fig}, the
smooth functions $e^{u}$, $e^{v}$, $e^{\phi_1/\sqrt2}$ and
$e^{\phi_2/\sqrt8}$ are plotted.

\subsection{Interpolating from AdS$_4 \times S^2$ to AdS$_6$}

      The AdS$_4\times S^2$ fixed-point arises for $m_1\, m_2 <0$ and
$\epsilon=+1$.  The Taylor expansion analysis indicates that there
are solutions with AdS$_4\times S^2$ in the asymptotic boundary as
well as at the near-horizon geometry. For a concrete example, we
choose $m_1=7$, $m_2=-5$, $g=1$ and a negative sign in
(\ref{ads4omega2}). The boundary solution behaves as
\bea e^{2u}&=& \ft94 (\ft{7}{36})^{3/4}\, y^{-2}\, (1 +
c\, y^n + \cdots)\,,\nn\\
e^{2v} &=& 3(\ft{343}{36})^{1/4}\, (1 -c\, y^n + \cdots)
\,,\nn\\
e^{\ft{\phi_1}{\sqrt2}} &=& \ft{1}{\sqrt{7}}\, (1 -
\ft{6(n+1)}{2n-11}c\,
y^n + \cdots)\,,\nn\\
e^{\ft{\phi_2}{\sqrt{8}}} &=& (\ft{7}{36})^{1/8}\, (1
-\ft{n+1}{2(n-3)}c\, y^n + \cdots)\,, \eea
where $n=1$ or $n=\ft12 (1+\sqrt{35})$, and $c$ is an arbitrary
integration constant.  Clearly, the solution runs from this
asymptotic region into a singularity in the bulk.

\begin{figure}
   \epsfxsize=4.0in \centerline{\epsffile{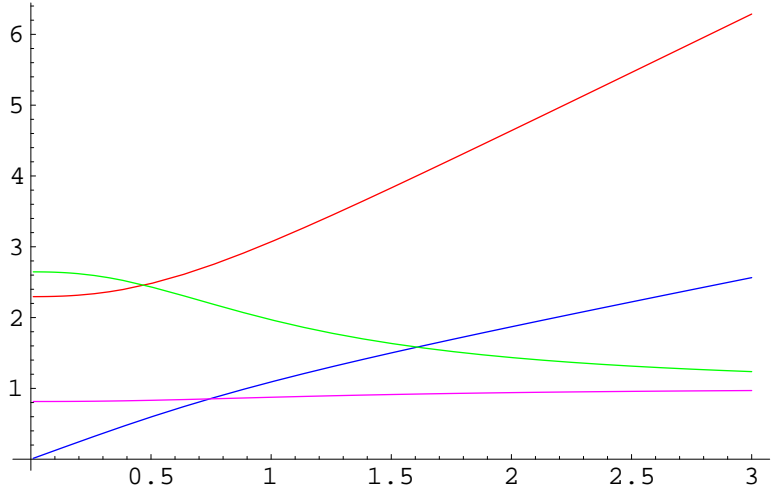}}
   \caption[FIG. \arabic{figure}.]{$e^u$ (blue), $e^{v}$
   (red), $e^{\phi_1/\sqrt2}$ (green) and $e^{\phi_2/\sqrt{8}}$
   (purple) for a smooth solution running from AdS$_4\times
   S^2$ at the horizon to an asymptotic AdS$_6$-type geometry.
   $m_1=7, m_2=-5$, $g=1$, and $c=-0.2$.} \label{2cd6s2fig}
\end{figure}

     In the solution whose near-horizon geometry is AdS$_4\times
S^2$, the Taylor expansion is given by
\bea
e^{2u} &=& (\ft{16384}{3087})^{1/4}\, (1 + c\, r^2 + \cdots)\,,\nn\\
e^{2v} &=& (\ft{3087}{4})^(1/4)\, (1 -\ft1{13}(19+2\sqrt{35})\,
c\, r^2 + \cdots)\,,\nn\\
e^{\ft{\phi_1}{\sqrt2}} &=& \sqrt7\, (1 + \ft{3}{65}(-25 +
9\sqrt{35})\, c\, r^2 + \cdots)\,,\nn\\
e^{\ft{\phi_2}{\sqrt8}} &=& (\ft{7}{36})^{1/8}\, (1 + \ft12
(-3 + 2 \sqrt{7/5})\, c\, r^2 + \cdots)\,,
\eea
where $c$ is an integration constant and $r$ is defined by $d\rho
=e^{-u}\, dr$.  We can use the above Taylor expansion as the
initial conditions for numerical calculation.  The plots of the
above functions are presented in Fig.~\ref{2cd6s2fig}, which
clearly show that the solution runs smoothly to the asymptotic
AdS$_6$-type solution given by (\ref{2cd6asym}).

\section{Three-charge string in $D=5$}

Let us now consider the minimal gauged supergravity in $D=5$
coupled to two vector multiplets. The Lagrangian is given by
\be e^{-1}{\cal L}_5=\hat R - \ft12 (\del \phi_1)^2 -\ft12 (\del
\phi_2)^2 - \fft14\sum_{i=1}^3 X_i^{-2} (\hat F_\2^i)^2 - \hat V +
e^{-1}\, \ft14 \epsilon^{\mu\nu\rho\sigma\lambda}\, \hat
F^1_{\mu\nu}\,\hat F^2_{\rho\sigma}\, \hat A^3_\lambda\,, \ee
with the scalar potential
\be
\hat V=-4g^2 \sum_{i=1}^3
         X_i^{-1}\,.\label{stdscalarpot}
\ee
The quantities $X_i$
are given by
\bea
&&X_i=e^{\ft12\vec a_i\cdot\vec \phi}\,,\nn\\
&&\vec a_1=(\sqrt2\,, \fft{2}{\sqrt6})\,, \qquad
\vec a_2=(-\sqrt2\,,\fft2{\sqrt6})\,,\qquad \vec
a_3=(0,- \,\fft4{\sqrt6})\,.
\eea
The scalar potential $\hat V$ in (\ref{stdscalarpot}) can be also
expressed in terms of the superpotential $\hat W$, given by
\be
\hat W=\sqrt2\, g\, \sum_i X_i\,.
\ee

We now reduce the theory on $S^2$ with the previous metric ansatz
specializing on $D=5$. The ansatz for the three $U(1)$ 2-form
field strengths is given by
\be
F_\2^i= \epsilon\,  m_i\, \lambda^{-2}\, \Omega_\2\equiv
\ft12 \epsilon\, q_i\, g^{-1}\, \Omega_\2\,.
\ee
The resulting $D=3$ scalar potential is
\be
V=\ft12 \epsilon^2 (\sum_i m_i^2\, X_i^{-2})\,
e^{4\alpha\, \varphi} -
2 \epsilon \lambda^2\, e^{3\alpha\, \varphi} +
\hat V\,e^{2\alpha\,\varphi}
\,,\label{stdd3pot}
\ee
and the corresponding superpotential is
\be
W=\ft{\epsilon}{\sqrt2}(\sum_im_i\,X^{-1}_i)\,
e^{2\alpha\, \varphi} +
\hat W\, e^{\alpha\, \varphi}\,,
\ee
provided that the following constraint is satisfied:
\be
\lambda^2=g\sum_i m_i\,.
\ee
Thus, the charge parameters $q_i$ satisfy $q_1 + q_2 + q_3=2$. The
first-order equations for the three-dimensional system have the
same form as (\ref{d7to5fo}) but now with $A'=-\fft{1}{\sqrt2}\,
W$. Lifting the equations, we obtain the first-order equations
describing the three-charge magnetic string in $D=5$:
\bea
\fft{d\vec \phi}{d\rho} &=&
\sqrt2\,\Big(-\fft{\epsilon}{2\sqrt2}\, (m_1\, \vec a_1\, X_1^{-1}
+ m_2\, \vec a_2\, X_2^{-1} + m_3\, \vec a_3\, X_3^{-1})\,e^{-2v}
+
\fft{d\hat W}{d\vec \phi}\Big)\,,\nn\\
\fft{dv}{d\rho} &=& -\fft{1}{3\sqrt2}\,
\Big(\sqrt2\, \epsilon\,(m_1\, X_1^{-1} +
m_2\, X_2^{-1} + m_3\, X_3^{-1})\, e^{-2v} + \hat W\Big)\,,\nn\\
\fft{du}{d\rho} &=& \fft{1}{3\sqrt2}\, \Big(
\fft{\epsilon}{\sqrt2}\,(m_1\, X_1^{-1} + m_2\, X_2^{-1} + m_3\,
X_3^{-1})\, e^{-2v} - \hat W\Big)\,. \label{5} \eea

A 3-charge string with constant scalars was found in
\cite{klemm2}. On the other hand, for $m_1=m_2=m_3$, it is
consistent that $\phi_1$ and $\phi_2$ vanish and this system
reduces to the $D=5$ system of section 2, with the solution given
by (\ref{sabra}) \cite{sabra1}. For $m_1=m_2$ we can consistently
set $\phi_1=0$ while keeping $\phi_2$. In particular, if
$m_1=m_2=0$, this system reduces to the $D=5$ system of section
3.1. After taking $g\rightarrow \sqrt{3/4}\, g$, an exact solution
is given by the formulae (\ref{gendmetricsol}) and (\ref{5H}).

For $m_1=m_2=m_3$, it has recently been found that one can
consistently set $\phi_1=\sqrt{3} \phi_2$, with an exact solution
given by \cite{recent}
\bea
ds_5^2&=&H^{-2/3}\Big[ (g\, r)^2 ({\rm e}^{\frac{\epsilon\,
m}{2g^3\, r^2}}dx_{\mu}^2+H^2 \lambda^{-2} d\Omega_2^2)+H^2
\frac{dr^2}{(g\, r)^2} \Big]\,,\nn\\
{\rm e}^{-\sqrt{6}\phi_2}&=&H\,, \label{new}
\eea
where we made the coordinate transformation
\be \frac{d\rho}{dr}=-\frac{{\rm e}^v}{(g\, r)^2}, \ee
and where
\be H=1-\frac{\epsilon\, m}{2g^3\, r^2}. \ee

For general $m_i$, we do not find analytical solutions and so it
is instructive to study the system with a numerical approach.

\subsection{Asymptotic boundary region}

We can express the metric in the form (\ref{expansion}) by making
the coordinate transformation $d\rho=dy\, {\rm e}^u$. The boundary
conditions are $u(y)\sim v(y)\sim -{\rm log}\, y$ and $\phi
\rightarrow 0$ for small $y$. The leading terms of the Taylor
expansion of the solution of (\ref{5}) are given by
\bea e^{2u} &=& \fft{1}{g^2\, y^2}\, ( 1 +
\fft{\epsilon\, (m_1 + m_2+m_3)\, g}{9 c^2}\, y^2 + \cdots)\,,\nn\\
e^{2v} &=& \fft{c^2}{g^2\, y^2}\, ( 1 -
\fft{7\epsilon\, (m_1+m_2+m_3)\, g}{18c^2}\, y^2 + \cdots)\,,\nn\\
e^{\ft{\phi_1}{\sqrt2}} &=& 1 - \fft{\epsilon\, (m_1 -
m_2)\,g}{2c^2}\, y^2\, {\rm log}\, y + \cdots\,,\\
e^{\fft{\phi_2}{\sqrt{6}}} &=& 1 - \fft{\epsilon\, (m_1 +
m_2-2m_3)\,g}{6c^2}\, y^2\, {\rm log}\, y + \cdots\,,\nn
\label{3cd5asym} \eea
where $c$ is an arbitrary integration constant. The
non-normalizable solutions for $\phi_1$ and $\phi_2$ imply that
the dual operators are turned on, provided that the values of
$m_i$ are such that the fields are non-vanishing \cite{nunez}.
Next, we examine how these solutions flow from the above boundary
into the bulk.

\subsection{AdS$_3\times H^2$ and AdS$_3\times S^2$}

For $\ep=\pm 1$, the fixed-point solution is given by
\bea e^{\sqrt2\, \phi_1} &=& \fft{m_1}{m_2}\,
\Big(\fft{m_3+m_2-m_1}{m_3 - m_2 + m_1}\Big)\,,\qquad e^{\sqrt6\,
\phi_2} = \fft{m_1\,m_2\,(m_3^2 - (m_1-m_2)^2)}{
m_3^2\,(m_1 + m_2 - m_3)^2}\,,\nn\\
e^{-2v}&=&-\epsilon\,g\,\Big(\fft{(m_1 + m_2 - m_3)(m_3^2 -
(m_1-m_2)^2)}{m_1^2\,m_2^2\,m_3^2}\Big)^{\ft13}\,,\nn\\
u&=& -g\, e^{\fft{\phi_2}{\sqrt6}}\, \Big(\cosh(\phi_1/\sqrt2) +
\ft12 e^{-\sqrt{\ft32}\,\phi_2}\Big)\, \rho
\equiv \fft{-2}{R_{\rm ads}}\, \rho\,.\label{ads3omega2}
\eea
This solution was discussed in detail in \cite{adsletter}. The reality
condition of the solution implies that when three vectors with the
magnitudes $|m_i|$ can form a triangle, $d\Omega_2^2$ should be the
$H^2$ metric. On the other hand, when they cannot form a triangle, the
metric should be that of $S^2$.\footnote{AdS$_3\times S^2$ solutions
were also recently found in \cite{recent} in a different
construction.}  If any of the $m_i$ vanish, there is no fixed-point
solution, except when one vanishes with the remaining two being equal.
The AdS$_3$ radius depends on $g$ and two of the three charge
parameters.

\subsection{Interpolating from AdS$_3 \times H^2$ to AdS$_5$}

As in the previous cases, AdS$_3\times H^2$ can occur either as an
asymptotic geometry or on the horizon. In order to demonstrate
this, we take $m_1=m_2=2$ and $m_3=3$. For this choice of charge
parameters, we can set $\phi_1=0$.  For simplicity, we also set
$g=1$.  As an asymptotic boundary geometry, the Taylor expansions
of the metric components and the scalars are
\bea
e^{2u}&=& \ft{8\, 2^{1/3}}{25}\, y^{-2} (1 + c\, y^n + \cdots)\,,\nn\\
e^{2v} &=& 2\, 2^{1/3}\, (1 -\ft12 c\,y^n
+\cdots)\,,\nn\\
e^{\ft{\phi_2}{\sqrt6}} &=& 2^{1/3}\, (1 + -\ft1{16}(11 +3\sqrt{17})
\, y^n + \cdots)\,,
\eea
where the coordinate $y$ is defined to be $d\rho=e^{u}\, dy$ and
$c$ is an arbitrary integration constant.  The constant $n$ can
take two values: $n=\ft15(-3+\sqrt{17})$.  Clearly, this solution
will flow into a singularity in the bulk.

\begin{figure}
   \epsfxsize=4.0in \centerline{\epsffile{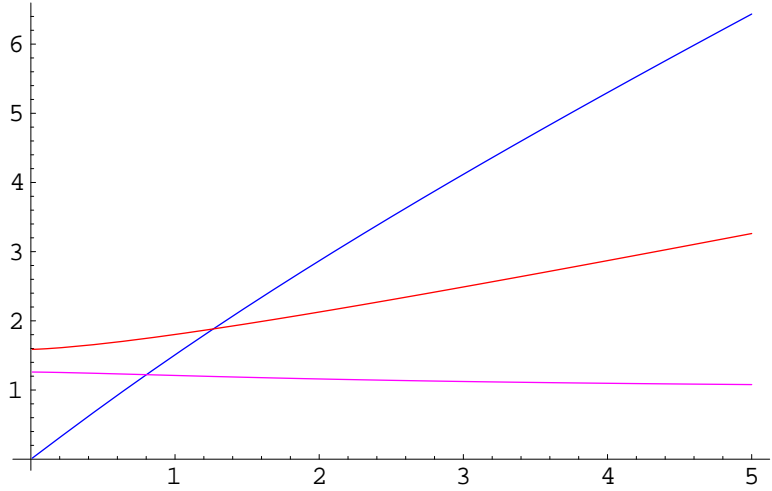}}
   \caption[FIG. \arabic{figure}.]{$e^u$ (blue), $e^{v}$
   (red) and $e^{\phi_2/\sqrt{6}}$
   (purple) for a smooth solution running from AdS$_3\times
   H^2$ at the horizon to an asymptotic AdS$_5$-type geometry.
   $m_1=m_2=2$, $m_3=3$, $g=1$, and $c=-0.05$.} \label{3cd5h2fig}
\end{figure}

     On the other hand, for the solution with AdS$_3\times H^2$ at
its horizon, we find the Taylor expansion of the next leading
order to be
\bea
e^{2u} &=& \ft{25}{8\, 2^{1/3}}\,r^2\, (1 + c\, r^2 + \cdots)\,,\nn\\
e^{2v} &=& 2\, 2^{1/3}\, (1 -\ft1{26}(33 + 10\sqrt{17})\, c\, r^2
+ \cdots)\,,\nn\\
e^{\ft{\phi_2}{\sqrt6}} &=& 2^{1/3}\, (1 + \ft{1}{208}\, (147 -
11\sqrt{17})\, r^2 + \cdots)\,,
\eea
where $c$ is an integration constant and $r$ is defined by
$d\rho=e^{-u}\, dr$.  We can use this as the initial conditions
for the numerical calculation. The resulting plots are presented
in Fig.~\ref{3cd5h2fig}), which show that the solution runs
smoothly from AdS$_3\times H^2$ at the horizon to the AdS$_5$-type
asymptotic behavior given by (\ref{3cd5asym}). The 3-charge string
with constant scalars also interpolates between AdS$_3\times H^2$
and the AdS$_5$-type geometry \cite{klemm2}.

\subsection{Interpolating from AdS$_3 \times S^2$ to AdS$_5$}

      AdS$_3\times S^2$ can also occur as an asymptotic geometry
for one solution or as a horizon geometry for another.  As a
concrete example, let us consider $m_1=m_2=-1$ and $m_3=4$, for
which case we have $\phi_1=0$.  For simplicity, we set $g=1$. The
boundary solution behaves as
\bea
e^{2u} &=& (\ft{9}{1024})^{1/3}\, y^{-2}\, (1 + c\, y^n + \cdots)\,,
\nn\\
e^{2v} &=& 6^{-1/3}\, (1 -\ft12 c\, y^n + \cdots)\,,\nn\\
e^{\ft{\phi_2}{\sqrt6}} &=& 6^{-1/3}\, (1 + \ft1{20} (11 + 3\sqrt{19})
\, c\, y^n + \cdots)\,,
\eea
where $n=\ft12(1 + \sqrt{19})$ and $c$ is an integration constant.
Clearly, the solution will encounter a singularity when it runs away
from this asymptotic region.

\begin{figure}
   \epsfxsize=4.0in \centerline{\epsffile{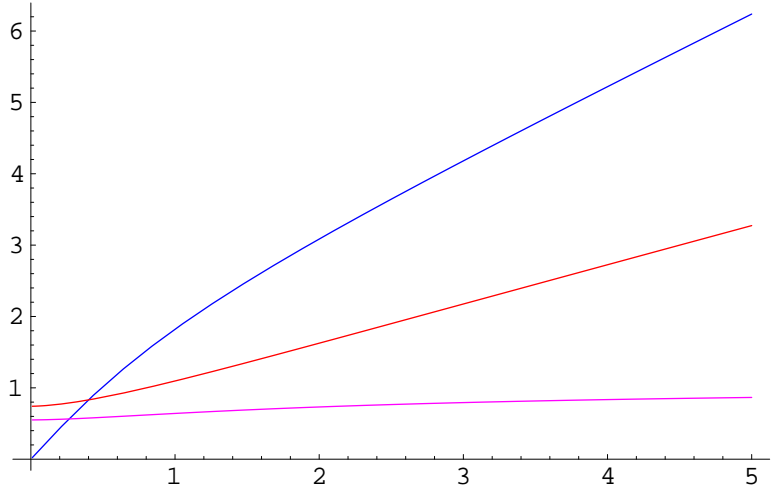}}
   \caption[FIG. \arabic{figure}.]{$e^u$ (blue), $e^{v}$
   (red) and $e^{\phi_2/\sqrt{6}}$
   (purple) for a smooth solution running from AdS$_3\times
   S^2$ at the horizon to an asymptotic AdS$_5$-type geometry.
   $m_1=m_2=-1$, $m_3=4$, $g=1$, and $c=-0.3$.} \label{3cd5s2fig}
\end{figure}

      For the solution with AdS$_3\times S^2$ at its horizon, the
near-horizon behavior is
\bea
e^{2u} &=& (\ft{1024}{9})^{1/3}\, (1 + c\, r^n + \cdots)\,,\nn\\
e^{2v} &=& 6^{-1/3}\, (1 -\ft1{10}(11 + 2\sqrt{19})\, c
\, r^n + \cdots)\,,\nn\\
e^{\ft{\phi_2}{\sqrt6}} &=& 6^{-1/3}\, (1 + \ft1{100}(7 -11
\sqrt{19})\, c\, r^n + \cdots)\,,
\eea
where $n=\ft12 (-1 + \sqrt{19})$ and $c$ is an integration
constant. We can use this as the initial conditions for a
numerical calculation and the results are plotted in
Fig.~\ref{3cd5s2fig}. This shows that the solution runs smoothly
from AdS$_3\times S^2$ at the horizon to the AdS$_5$-type
asymptotic geometry given by (\ref{3cd5asym}).

\section{Four-charge black hole in $D=4$}

Let us now consider the $U(1)^4$ gauged $N=2$ supergravity in four
dimensions. The Lagrangian is given by
\be e^{-1}{\cal L}_4=\hat R - \ft12 (\del \phi_1)^2 -\ft12 (\del
\phi_2)^2 -\ft12 (\del \phi_3)^2 \fft14\sum_{i=1}^4 X_i^{-2} (\hat
F_\2^i)^2 - \hat V \,, \ee
with the scalar potential
\be \hat V=-4g^2 \sum_{i<j} X_i\, X_j\,. \label{d4V} \ee
The quantities $X_i$ are given by
\bea
&&X_i=e^{\ft12\vec a_i\cdot\vec \phi}\,,\nn\\
&&\vec a_1=(1, 1, 1),\quad \vec a_2=(1, -1, -1),\quad \vec
a_3=(-1, 1, -1),\quad \vec a_4=(-1, -1, 1). \eea
The scalar potential $\hat V$ in (\ref{d4V}) can be also expressed
in terms of the superpotential $\hat W$, where
\be \hat W=\sqrt2\, g\, \sum_{i=1}^4 X_i\,. \ee
We use the metric ansatz (\ref{metricansatz}) and take the four
$U(1)$ 2-form field strengths to be
\be F_\2^i= \epsilon  m_i\, \lambda^{-2}\, \Omega_\2
\equiv \ft12\epsilon\,q_i\,g^{-1}\, \Omega_\2\,. \ee
A solution will be supersymmetric provided that the following
constraint is satisfied:
\be \lambda^2=g\sum_i m_i\,, \ee
and hence $q_1 + q_2 + q_3 + q_4=2$.  The equations of motion are
given by
\bea \fft{d\vec \phi}{d\rho} &=&
\sqrt2\,\Big(-\fft{\epsilon}{2\sqrt2}\, \sum_{i=1}^4 m_i\, \vec
a_i\, X_i^{-1}\, \,e^{-2v} +
\fft{d\hat W}{d\vec \phi}\Big)\,,\nn\\
\fft{dv}{d\rho} &=& -\fft{1}{2\sqrt2}\,
\Big(\fft{\epsilon}{\sqrt{2}}\,
\sum_{i=1}^4 m_i\, X_i^{-1}\, e^{-2v} + \hat W\Big)\,,\nn\\
\fft{du}{d\rho} &=& \fft{1}{2\sqrt2}\, \Big(
\fft{\epsilon}{\sqrt2}\, \sum_{i=1}^4 m_i\, X_i^{-1}\, e^{-2v} -
\hat W\Big)\,. \label{4} \eea

For $m_1=m_2=m_3=m_4$, it is consistent that $\phi_1$ and $\phi_2$
vanish and this system reduces to the $D=4$ system of section 2,
with the solution given by (\ref{sabra}) \cite{sabra1}. For
general $m_i$, we do not find analytical solutions and it is
instructive to study the system using the numerical approach.

\subsection{Asymptotic boundary region}

We can express the metric in the form (\ref{expansion}) by making
the coordinate transformation $d\rho=dy\, {\rm e}^u$. We have the
boundary conditions $u(y)\sim v(y)\sim -{\rm log}\, y$ and $\phi
\rightarrow 0$ for small $y$. The leading terms of the Taylor
expansion of the solution of (\ref{4}) are given by
\bea e^{2u} &=& \fft{1}{4g^2\, y^2}\, ( 1 +
\fft{\epsilon\, \sum_{i=1}^4 m_i\, g}{3 c^2}\, y^2 + \cdots)\,,\nn\\
e^{2v} &=& \fft{c^2}{4g^2\, y^2}\, ( 1 -
\fft{2\epsilon\, \sum_{i=1}^4 m_i\, g}{3c^2}\, y^2 + \cdots)\,,\nn\\
e^{\ft{1}{2}{\vec \phi}} &=& 1 - \fft{\epsilon\, \sum_{i=1}^4
{\vec a}_i\, m_i\,g}{2c^2}\, y^2 + \cdots\,,\label{4cd4asym} \eea
where $c$ is an arbitrary integration constant. As in the cases of
$D=7$ and $6$, the normalizable solutions for $\phi_1$ and
$\phi_2$ imply that the dual operators are not turned on. Next, we
examine how these solutions flow from the above boundary into the
bulk.

\subsection{AdS$_2\times H^2$ and AdS$_2\times S^2$}

For $\epsilon=\pm 1$, we have not obtained the general solution
for arbitrary $m_i$. However, a class of special solutions results
from setting $m_2=m_3=m_4$ \cite{gauntlett4}.  This enables one to
consistently set $\phi_1=\phi_2=\phi_3\equiv \phi$. For this
truncation, fixed-point solutions for $\epsilon = \pm 1$ are given
by
\bea {\rm e}^{2 \phi} &=& \frac{3m_2-m_1\pm
\sqrt{(m_1-m_2)(m_1-9m_2)}}{2m_2}\,,\qquad {\rm e}^{-2v}= 4g\,
\epsilon\, \frac{{\rm sinh}\, \phi}{m_1\, {\rm
e}^{-2\phi}-m_2}\,,\nn\\
u&=& \ft12g\,\Big( \frac{2(m_1\, {\rm e}^{-\frac{3}{2}\phi}+3m_2\,
{\rm e}^{\frac{1}{2}\phi})}{m_2-m_1\, {\rm e}^{-2\phi}}\, {\rm
sinh}\, \phi+{\rm e}^{\frac{3}{2}\phi}+3{\rm
e}^{-\frac{1}{2}\phi}\Big)\, \rho=
\fft{2}{R_{\rm ads}}\, \rho\,.\label{ads2omega2} \eea
This solution was discussed in detail in \cite{adsletter,
gauntlett4}. The reality condition of the solution implies that for
$\ep=-1$, corresponding to $H^2$, we must have either $m_2 >0$ and
$0<m_1\le m_2$ or $m_2<0$ and $m_2\le m_1 < -3m_2$. For $\ep=1$,
corresponding to $S^2$, we must have $m_2\le 0$ and $m_1>-3m_2$.  In
general, the AdS$_2$ radius depends on $g$ and three of the four
charge parameters.

\subsection{Interpolating from AdS$_2 \times H^2$ to AdS$_4$}

As in the previous cases, AdS$_2\times H^2$ can occur at the
asymptotic boundary of one solution and the horizon geometry of
another.  This can be demonstrated using Taylor expansion. For a
concrete example, let us consider $m_1=2$ and $m_2=m_3=m_4=3$.  As
shown in the previous subsection, we have $\phi_i=\phi$.  Near the
asymptotic boundary, the solution behaves as
\be e^{2u}= \ft{\sqrt2}{25}\,y^{-2}\, (1 + c\, y^{2/5} +
\cdots)\,,\quad e^{2v}=\sqrt2\, (1 + \cdots)\,,\quad e^{\ft12\phi}
= 2^{1/4}\, (1-\ft76\, c\, y^{2/5} + \cdots). \ee
Clearly this solution will run into a singularity in the bulk.

\begin{figure}
   \epsfxsize=4.0in \centerline{\epsffile{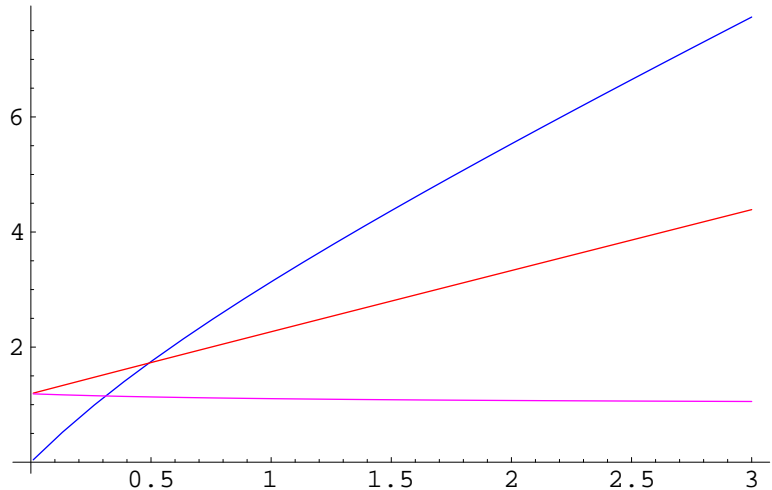}}
   \caption[FIG. \arabic{figure}.]{$e^u$ (blue), $e^{v}$
   (red), and $e^{\phi_2/2}$
   (purple) for a smooth solution running from AdS$_2\times
   H^2$ at the horizon to an asymptotic AdS$_4$-type geometry.
   $m_1=2$, $m_2=m_3=m_4=3$, $g=1$, and $c=-0.5$.} \label{4cd4h2fig}
\end{figure}

     On the other hand, for the solution with the AdS$_2\times H^2$
horizon, the near-horizon behavior is given by
\be
e^{2u}=\ft{25}{\sqrt2}\, r^2\, (1 + c\, r + \cdots)\,,\quad
e^{2v}= \sqrt2\, (1 -\ft{35}{19}\,c\, r+ \cdots)\,,\quad
e^{\ft12\phi} = 2^{1/4}\, (1 + \ft{5}{38}\, c\, r
+\cdots)\,,
\ee
where $c$ is an integration constant.  Using this as initial data
for numerical calculation, we plot the result in
Fig.~\ref{4cd4h2fig}. Thus, we see that the solution runs from
AdS$_2\times H^2$ at the horizon to the AdS$_4$-like geometry
(\ref{4cd4asym}) in the asymptotic region. A black hole solution
which interpolates between AdS$_2\times H^2$ and the AdS$_4$-type
geometry was first found in \cite{klemm1}.

\subsection{Interpolating from AdS$_2 \times S^2$ to AdS$_4$}

      Analogously, AdS$_2\times S^2$ can also either occur at the
horizon region of a solution and at the asymptotic region of
another. As an example, let us consider $m_1=7 $ and
$m_2=m_3=m_4=-2$, which implies that $\phi_i=\phi$.  For the
solution with AdS$_2\times S^2$ in its asymptotic region, the
behavior can be studied using Taylor expansion:
\be
e^{2u}=\ft{\sqrt7}{100}\, (1 + c\, r^{6/5} + \cdots)\,,\quad
e^{2v}=\ft{\sqrt7}{4}\, (1 + \cdots)\,,\quad
e^{\ft12\phi}=7^{1/4}\, (1 - \ft{11}{18}\, c\, r^{6/5} +
\cdots)\,,
\ee
where $c$ is an arbitrary integration constant.  Clearly, the flow
of this solution from the boundary leads to a singularity in the
bulk.

\begin{figure}
   \epsfxsize=4.0in \centerline{\epsffile{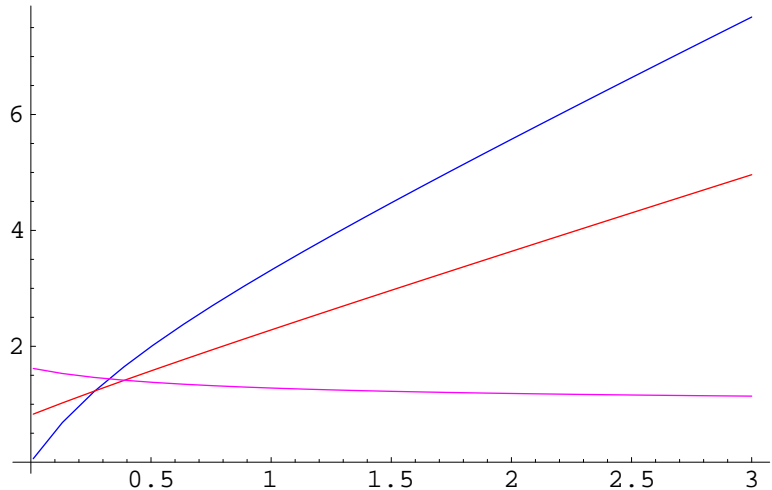}}
   \caption[FIG. \arabic{figure}.]{$e^u$ (blue), $e^{v}$
   (red), and $e^{\phi_2/2}$
   (purple) for a smooth solution running from AdS$_2\times
   S^2$ at the horizon to an asymptotic AdS$_4$-type geometry.
   $m_1=7$, $m_2=m_3=m_4=-2$, $g=1$, and $c=-1.5$.} \label{4cd4s2fig}
\end{figure}

      For the solution with AdS$_2\times S^2$ at its horizon,
the near-horizon behavior is given by
\be
e^{2u}=\ft{100}{\sqrt7}\, r^2\, (1 + c\, r + \cdots)\,,\quad
e^{2v}=\ft{\sqrt7}{4} (1 -\ft{55}{41}\, c\, r + \cdots)\,,\quad
e^{\ft12\phi} =7^{1/4}\, (1 +\fft{15}{82}\, c\, r +\cdots)\,,
\ee
where $c$ is an integration constant.  Using this as initial data, we
perform numerical calculation.  The plots are presented in
Fig.~\ref{4cd4s2fig}.  From the plots, we see that the solution runs
smoothly from the AdS$_2\times S^2$ horizon to the AdS$_4$-like
asymptotic region given in (\ref{4cd4asym}).

\section{String and M-theory origins}

          Gauged supergravities in $D=4,5,6$ and $7$ can be obtained
from consistent sphere reductions of M-theory, type IIB or massive
type IIA supergravities.  Thus, it is straightforward to lift our
solutions back to higher dimensions and study their properties.  A
particularly interesting class of solutions which we obtained are
those that smoothly interpolate between AdS$_{D-2}\times \Omega_2$
at the horizon and AdS$_{D}$-type spacetime in the asymptotic
region. When lifted to higher dimensions, the asymptotic region
becomes a direct product of AdS$_{D}$ with a relevant sphere
\footnote{Except in the case of massive type IIA supergravity, for
which the AdS$_6$ embedding involves a singular warp factor
\cite{oz,massive}.}.  On the other hand, the horizon region becomes a
warped product of AdS$_{D-2}$ with an internal metric that is an $S^p$
bundle over $S^2$ for $p=4,5$ and 7.  We shall look at these cases in
detail.  Some of the results have been reported in a recent letter
\cite{adsletter}.

\subsection{$D=7$ solutions embedded in M-theory}

      It is straightforward to lift the AdS$_5\times S^2$ and
AdS$_5\times H^2$ solutions given by (\ref{ads5s2}) to $D=11$ by
using the ansatz obtained in \cite{tenauthors}.  Since the
solutions for general $m_i$ are rather complicated to present, we
only consider a representative example of AdS$_5\times S^2$ with
$m_1=5g$ and $m_2=-3g$.  The M-theory metric is given by
\bea ds_{11}^2 &=& \Delta^{\ft13}\,\Big[ds_{\rm{AdS_5}}^2 +
\fft{1}{g^2\, c}\, \Big\{ \fft{1}{4c}\,
(d\theta^2 + \sin^2\theta\, d\varphi^2)\nn\\
&&+ \fft{1}{\Delta}\,\Big(\ft14\, d\mu_0^2 +
\ft15(d\mu_1^2 + \mu_1^2\,
(d\phi_1 -\ft52 \cos\theta\, d\varphi)^2)\nn\\
&& + d\mu_2^2 + \mu_2^2\, (d\phi_2 + \ft32\cos\theta\, d\varphi)^2
\Big) \Big\}\Big]\,, \eea
where $c=10^{-2/5}$ and $\mu_i$ are spherical coordinates which
satisfy $\mu_0^2 + \mu_1^2 + \mu_2^2=1$. The warp factor $\Delta$,
which is not to be confused with the dilaton coupling parameter in
section 3, is given by
\be \Delta=c\, (4\mu_0^2 + 5 \mu_1^2 + \mu_2^2)>0\,. \ee
The AdS$_5$ metric is given by
\be ds_{\rm{AdS_5}}^2 = e^{-\ft{2\rho}{R}}\, dx^\mu\, dx_\mu +
d\rho^2\,, \ee
where the AdS radius is $R=\fft{1}{2c\,g}$. Note that, for all
lifted solutions, $g$ is defined as in sections 4-7.

The case of $m_1=m_2$ provides an example of AdS$_5\times H^2$
which is easily embedded in M-theory. The corresponding
eleven-dimensional metric is given by
\bea ds_{11}^2 &=& \Delta^{\ft13}\,\Big[ds_{\rm{AdS_5}}^2 +
(\ft34)^{3/5} \fft{1}{g^2}\, \Big\{ d\theta^2 + \sin^2\theta\,
d\varphi^2 +d\psi_1^2\nn\\ & & + (\ft34)^{-4/5} \ft{1}{\Delta}\,
{\rm cos}^2\, \psi_1\, \Big( d\psi_2^2+{\rm cos}^2 \psi_2\,
(d\phi_1- \ft12 {\rm cosh}\, \theta\, d\varphi )^2\nn\\ & & +{\rm
sin}^2 \psi_2\, (d\phi_2-\ft12 {\rm cosh}\, \theta\, d\varphi)^2
\Big) \Big\} \Big]\,. \eea
The warp factor $\Delta$ is given by
\be \Delta=(\ft34)^{-4/5}{\rm sin}^2\, \psi_1+(\ft34)^{1/5}{\rm
cos}^2\, \psi_1 >0\,, \ee
and the AdS radius is $R=(\ft34)^{4/5}\, \ft2g$.

     In the above geometries, the internal space of the $D=11$ metric
can be viewed as an $S^4$ bundle over $S^2$ or $H^2$, with two
diagonal $U(1)$ bundles \footnote{Supersymmetric AdS solutions of
M-theory with internal spaces which can be viewed as $S^4$ bundles
over various spaces have also been discussed in
\cite{gauntlett1,gauntlett2,gauntlett3}.}. In general, the
internal metric can be labelled by the two diagonal monopole
charges $(q_1, q_2)=(\fft{2m_1}{m_1+m_2}, \fft{2m_2}{m_1 + m2})$.
In the specific example of AdS$_5\times S^2$ above,
$(q_1,q_2)=(5,-3)$ and the solution is smooth everywhere. For
general $(m_1,m_2)$, the metric does not have a power-law
singularity. However, it could have a conical orbifold
singularity, which is absent only if $(q_1,q_2)$ are integers.
Since $q_i$ satisfy the constraint $q_1 + q_2=2$, it follows that
they are either both even or both odd integers. In the even case,
the bundle is topologically trivial, whilst it is twisted for the
odd case. The internal space can be regarded as having a
generalized holonomy group \cite{duffliu}, since it is not Ricci
flat and involves a form field.

     The solution preserves ${\cal N}=2$ supersymmetry; it is a
supergravity dual to an ${\cal N}=1$, $D=4$ superconformal field
theory on the boundary of AdS$_5$. This provides a concrete framework
to study AdS$_5$/CFT$_4$ from the point of view of M-theory.

     The seven-dimensional single-charge 3-brane given by
(\ref{singlec}) and (\ref{d7h}) can also be lifted to M-theory.  The
corresponding eleven-dimensional metric is given by
\be ds_{11}^2= \Delta^{1/3} ds_7^2+\frac{1}{g^2 \Delta^{2/3}}
\Big[ H^{-1/5} \Delta\, d\theta^2+H^{2/5} {\rm cos}^2 \theta\,
d{\td \Omega}_2^2+H^{-3/5} {\rm sin}^2 \theta\, (d\phi_1+A_\1)^2
\Big], \ee
where
\be \Delta=H^{3/5}{\rm sin}^2 \theta+H^{-2/5}{\rm cos}^2 \theta >0
\,, \qquad dA_\1=\Omega_\2\,. \ee

\subsection{$D=6$ solutions embedded in massive IIA theory}

         Using the reduction ansatz in \cite{gubser,massive}, it is
straightforward to lift the AdS$_4\times S^2$ and AdS$_4\times
H^2$ solutions given by (\ref{ads4omega2}) up to $D=10$ massive
IIA supergravity. The metric is given by
\bea ds_{10}^2 &=& \mu_0^{\ft1{12}}\, X_0^{\ft18}\, (X1\,
X_2)^{\ft14} \,\Delta^{\ft38}\, \Big[ds_6^2 + g^{-2}\,
\Delta^{-1}\,
\Big(X_0^{-1}\, d\mu_0^2\\
&&+X_1^{-1}\, (d\mu_1^2 + \mu_1^2\,(d\varphi_1 + g\,A_\1^2)^2) +
X_2^{-1}\, (d\mu_2^2 + \mu_2^2\,(d\varphi_2 +
g\,A_\2^1)^2)\Big)\Big] \,,\nn \eea
where $\Delta=\sum_{\alpha=0}^2 X_\a\,\mu_\alpha^2>0$ and $\m_0^2
+ \mu_1^2 + \mu_2^2=1$.  Thus, the $D=10$ metric is a warped
product of AdS$_4$ with an internal six-metric, which is an $S^4$
bundle over $S^2$ or $H^2$, depending on the charge parameter $p$
according to the rule (\ref{d6rule}).

    As an example of a supersymmetric, though singular,
compactification of AdS$_4$ from massive IIA theory, we can take
$m_1=7g$, $m_2=-5g$ and a negative sign in (\ref{ads4omega2}).  This
gives $X_0=6c$, $X_1=7c$ and $X_2=c$, where $c=6^{-1/4}\,
7^{-3/8}$. Also, $A_\1^1=-\frac{7}{2g}{\rm cos}\, \theta\, d\varphi$,
$A_\1^2=\frac{5}{2g}{\rm cos}\, \theta\, d\varphi$ and the radius of
AdS$_4$ is given by $R=1/(4c\, g)$. On the other hand, $m_1=m_2$
provides an embedding of AdS$_4\times H^2$ in massive IIA theory.

The six-dimensional single-charge membrane given by
(\ref{singlec}) and (\ref{d6h}) can also be lifted to massive IIA
theory. The corresponding ten-dimensional metric is given by
$$
ds_{10}^2= ({\rm cos}\, \theta\, {\rm cos}\, \psi
)^{1/12}H^{1/64}\Big( \Delta^{3/8} ds_6^2+\frac{1}{g^2
\Delta^{5/8}} \Big[ H^{-1/4} \Delta\, d\theta^2+
$$
\be H^{3/8} {\rm cos}^2 \theta\, d{\td \Omega}_2^2+H^{-5/8} {\rm
sin}^2 \theta\, (d\phi_1+A_\1)^2 \Big] \Big), \ee
where
\be \Delta=H^{5/8}{\rm sin}^2 \theta+H^{-3/8}{\rm cos}^2 \theta >0
\,,\qquad dA_\1=\Omega_\2\,. \ee
Note that the overall warping factor depends on two internal
coordinates.

\subsection{$D=5$ solutions embedded in IIB theory}

        The AdS$_3\times S^2$ and AdS$_3\times H^2$ solutions given by
(\ref{ads3omega2}) can be lifted to ten-dimensional IIB theory
with the reduction ansatz obtained in \cite{tenauthors}.  Since
the solution with general $m_i$ is complicated to present, we
consider a simpler case with $m_2=m_1$. The ten-dimensional metric
is
\bea ds_{10}^2&=& \sqrt{\Delta}\Big\{ ds_{\rm{AdS_3}}^2
+\epsilon\,g^{-2}
(\fft{m_1}{m_3-2m_1})^{1/3}\, (\ft12 q_1\, d\Omega_2^2 + d\theta^2)\nn\\
&&+ g^{-2}\,\Delta^{-1}\,\Big[ c^{-1/3}\, \cos^2\theta\,\Big
(d\psi^2 + \sin^2\psi\, (d\varphi_1 + \ft12 q_1\, A_\1)^2\nn\\
&& +\cos^2\psi\, (d\varphi_2 + \ft12 q_1\, A_\1)^2\Big) +
c^{2/3}\, \sin^2\theta\, (d\varphi_3 + \ft12 q_3\,
A_\1)^2\Big]\Big\}\,, \eea
where
\bea &&c=\Big|\fft{m_1}{2m_1-m_3}\Big|\,,\qquad \Delta=c^{1/3}\,
\cos^2\theta
+ c^{-2/3} \sin^2\theta>0\,,\qquad dA_\1=\Omega_\2\,,\nn\\
&&ds_{\rm{AdS_3}}^2 = e^{-\fft{2\rho}{R}}\, (-dt^2 + dx^2) +
d\rho^2\,, \qquad R=\Big|\fft{2m_1}{g\,(4m_1-m_3)\, c^{1/3}}\Big|
\,. \eea
We have introduced the charge parameters $q_i=2 m_i/(m_1 + m_2 +
m_3)$, and hence they satisfy the constraint $q_1 + q_2 + q_3=2$.
In the above solution, if $|m_3|< 2|m_1|$, we should have
$\epsilon=-1$, corresponding to $H^2$; if $|m_3|> 2|m_1|$, we
should have $\epsilon=1$, corresponding to $S^2$.  In general, the
internal metric is an $S^5$ bundle over $S^2$ or $H^2$, depending
the values of the $q_i$ according to the above rules.

It is especially simple to lift the five-dimensional
equal-three-charge string given by (\ref{sabra}), due to the
absence of scalars. The corresponding ten-dimensional metric is
given by
$$
ds_{10}^2=ds_5^2+\frac{1}{g^2} \Big[ d\theta^2+{\rm cos}^2 \theta
\Big( d\psi^2+{\rm sin}^2 \psi (d\phi_1+\frac{1}{3}A_\1)^2
$$
\be +{\rm cos}^2 \psi (d\phi_2+\frac{1}{3}A_\1)^2 \Big)+{\rm
sin}^2 \theta (d\phi_3+\frac{1}{3}A_\1)^2 \Big], \ee
where $dA_\1=\Omega_\2$.

For the five-dimensional three-equal-charged string with a
nontrivial dilaton, given by (\ref{new}), the corresponding
ten-dimensional metric is
$$
ds_{10}^2=\sqrt{\Delta}ds_5^2+\frac{1}{g^2 \sqrt{\Delta}} \Big[
\Delta H^{1/3}d\theta^2+H^{2/3}{\rm sin}^2
\theta(d\phi_1+\frac{1}{3} A_\1)^2+
$$
\be H^{-1/3}{\rm cos}^2 \theta \Big( d\psi^2+{\rm sin}^2 \psi
(d\phi_2+\frac{1}{3} A_\1)^2+{\rm cos}^2 \psi (d\phi_3+\frac{1}{3}
A_\1)^2 \Big) \Big], \ee
where
\be \Delta=H^{1/3}{\rm cos}^2 \theta+H^{-2/3}{\rm sin}^2 \theta >0
\,, \qquad dA_\1=\Omega_\2\,. \label{delta} \ee

For five-dimensional single-charge string given by (\ref{singlec})
and (\ref{5H}), the corresponding metric is
\be ds_{10}^2=\sqrt{\Delta}ds_5^2+\frac{1}{g^2 \sqrt{\Delta}}
\Big[ H^{1/3} \Delta\, d\theta^2+H^{-1/3}{\rm cos}^2 \theta\,
d\Omega_3^2 +H^{2/3}{\rm sin}^2 \theta (d\phi_3+A_\1)^2 \Big], \ee
where $\Delta$ and $A_\1$ are given by (\ref{delta}).

\subsection{$D=4$ solutions embedded in M-theory}

        We use the reduction ansatz obtained in \cite{tenauthors} to
lift the AdS$_2\times S^2$ and AdS$_2\times H^2$ solutions given
by (\ref{ads2omega2}) back to $D=11$, with the metric
\bea ds_{11}^2 &=& \Delta^{2/3}\Big\{ ds_{\rm{AdS_2}}^2
+\frac{{\rm
e}^{2v}}{(m_1+3m_2)g}d\Omega_2^2\nn\\
& &+\frac{1}{g^2\, \Delta} \Big[ {\rm e}^{-\frac{3}{2}\phi}\Big(
d\mu_1^2+\mu_1^2(d\phi_1+\frac{\epsilon\,
m_1}{m_1+3m_2} A_\1)^2 \Big)\nn\\
& &+{\rm e}^{\frac{1}{2}\phi} \sum_{i=1}^3 \Big( d\mu_i^2+\mu_i^2
(d\phi_i +\frac{\epsilon\, m_2}{m_1+3m_2} A_\1)^2 \Big) \Big]
\Big\}\,, \eea
where
\bea \Delta &=& ({\rm e}^{\frac{3}{2}\phi}-{\rm
e}^{-\frac{1}{2}\phi})\mu_1^2+{\rm e}^{-\frac{1}{2}\phi}>0\,,
\quad dA_\1=\Omega_\2\,,\quad ds_{\rm{AdS_2}}^2 =
-e^{-\fft{2\rho}{R}}\, dt^2 + d\rho^2\,,
\nn\\
R &=& \frac{2}{g} \Big[ \frac{2(m_1\, {\rm
e}^{-\frac{3}{2}\phi}+3m_2\, {\rm e}^{\frac{1}{2}\phi})}{m_2-m_1\,
{\rm e}^{-2\phi}}\, {\rm sinh}\, \phi+{\rm
e}^{\frac{3}{2}\phi}+3{\rm e}^{-\frac{1}{2}\phi}\Big]^{-1}\,. \eea
In general, the nine-dimensional internal metric is an $S^7$
bundle over $S^2$ or $H^2$, depending the values of the $m_i$.
This is an especially interesting example of a space with
generalized holonomy group, since nine-dimensional Ricci-flat
manifolds do not have an irreducible special holonomy group.

The four-dimensional single-charge black hole given by
(\ref{gendmetricsol}) and (\ref{d4h}) can also be embedded in
M-theory, with the corresponding eleven-dimensional metric
$$ ds_{11}^2= \Delta^{2/3} ds_4^2+\frac{H^{1/4}}{g^2 \Delta^{1/3}}
\Big[ \Big( 1+(H^{-1/2}-1){\rm sin}^2 \theta\, {\rm cos}^2 \varphi
\Big) d\theta^2+
$$
$$
{\rm cos}^2 \theta\, \Big( 1+(H^{-1/2}-1){\rm sin}^2 \varphi \Big)
d\varphi^2+\frac{1}{2}(H^{-1/2}-1){\rm sin}(2\theta)\, {\rm
sin}(2\varphi)\, d\theta\, d\varphi+
$$
\be H^{-1/2}{\rm cos}^2 \theta\, {\rm cos}^2 \varphi\,
d\Omega_3^2+{\rm sin}^2 \theta\, (d\phi_1+A_\1)^2+{\rm cos}^2
\theta\, {\rm sin}^2 \varphi\, d\phi_2^2 \Big], \ee
where
\be \Delta=H^{-1/4}({\rm sin}^2 \theta+{\rm cos}^2 \theta\, {\rm
sin}^2 \varphi)+H^{1/4}{\rm cos}^2 \theta\, {\rm cos}^2 \varphi >0
\,, \quad dA_\1=\Omega_\2\,. \ee

\section{Conclusions}

      We have investigated a large class of supersymmetric
magnetic brane solutions supported by $U(1)$ gauge fields in
gauged supergravities of dimensions $4\le D\le 7$.  We have
obtained first-order equations by using a superpotential approach.
These equations admit stationary AdS$_{D-2}\times \Omega^2$
solutions, where $\Omega^2$ can be $S^2$ or $H^2$ depending on the
values of the $U(1)$-charges.  The $U(1)$-charges $q_i$ satisfy
the condition
\be
q_1 + q_2 + \cdots + q_n=2\,.
\ee
For $D=7$ and $6$, a maximum of two $U(1)$ charges are allowed and the
rule for having $S^2$ or $H^2$ can be clearly summarized by
(\ref{s2h2rule}).  For $D=5$ and $4$, the maximum number of charges
can be 3 and 4, respectively, and the complete rule remains to be
specified, although we have obtained explicit examples of $S^2$ and
$H^2$.  Such constraints show that magnetic branes in AdS gauged
supergravity are rather different from those in ungauged
supergravities since, for the latter case, charges are unconstrained
integration constants.  The radius of AdS$_{D-2}$ depends on the
$D$-dimensional cosmological constant and $(n-1)$ charge parameters.

       The fraction of supersymmetry preserved by our brane solutions
is the same as for those in ungauged supergravities, and depends on
the number of $U(1)$ field strengths involved. Embedded in maximal
supersymmetric theories, the preserved fraction of the supersymmetry
is $2^{-N}$, for $N=1$, 2, or 3 field strengths. In the case of $N=4$,
the introduction of the fourth field strength does not break further
supersymmetry and hence the preserved fraction is $1/8$.

      Our first-order equations support two types of solutions that
run from the aforementioned stationary ones. The first type has a
geometry which approaches AdS$_{D-2}\times \Omega^2$ in the
asymptotic region and the solution runs into a singularity at
small distance. The second type, which is more interesting, has a
near-horizon geometry of AdS$_{D-2}\times \Omega^2$ and smoothly
runs to an AdS$_D$-type geometry (\ref{adslike}) in the asymptotic
region. The non-constant scalar fields are bounded. This implies
the existence of a large class of conformal field theories in
diverse dimensions whose renormalization group flows from the UV
conformal fixed point to the IR conformal fixed point. The fact
that, from the same set of equations, AdS$_{D-2}\times \Omega^2$
arises as the horizon geometry in one type of solution and the
asymptotic geometry in another, suggests that the stationary
solution lies at the inflection point of the modulus space.  Thus,
a phase transition can occur, determining whether the system runs
into a singularity or an AdS$_D$-type metric.

     The asymptotic AdS$_D$-type geometry of the second type of
solution has a boundary of Minkowski$_{D-3}\times \Omega^2$. The case
of $D=7$ arises as a vacuum solution of a six-dimensional gauged
supergravity constructed in \cite{sezgind6}.  This vacuum solution is
of particular interest since the $SU(2)$ Yang-Mills field can be
obtained from the internal $S^2$ without introducing a non-vanishing
cosmological constant.  Similar possibilities have also been observed
for $S^3$ and brane-world Kaluza-Klein reduction of string theory
\cite{liulupope}. However, unlike these present solutions, the
reduction discussed in \cite{liulupope} is non-supersymmetric and
singular. Of course, the asymptotic AdS$_D$-type geometry
(\ref{adslike}) is not an exact solution everywhere, since it runs to
AdS$_5\times S^2$ in the bulk.  We do not expect that there exists a
consistent Kaluza-Klein reduction of $D=7$ gauged supergravity to the
$D=6$ theory of \cite{sezgind6}. Rather, the latter theory should be
viewed as the effective theory only at the boundary of our
solutions. Since there are also solutions that smoothly run from
AdS$_5\times H^2$ to an AdS$_7$-type geometry with the boundary
Minkowski$_4\times H^2$, we expect that there exists an effective
gauged supergravity theory in $D=6$ that admits such a vacuum
solution.

\section*{ACKNOWLEDGMENTS}

        We are grateful to Gary Gibbons, Chris Pope and Ergin Sezgin
for useful discussions.

\end{document}